\documentclass{mn2e}

\usepackage{epsfig}

\usepackage{mathptm}

\title{Some estimates of the source counts at \textit{Planck Surveyor} frequencies, using the 9C survey at 15 GHz}
\author[ ]
       { E.M. Waldram, R.C. Bolton, G.G. Pooley, and J.M. Riley
\\
        Astrophysics Group, Cavendish Laboratory, Madingley Road, Cambridge, CB3 0HE}
\date{   }

\pagerange{\pageref{firstpage}--\pageref{lastpage}}
\pubyear{2004}

\begin{document}

\maketitle

\label{firstpage}

\begin{abstract}
We have used multi-frequency follow-up observations of a sample of extragalactic sources from the 9C~survey at 15~GHz to make deductions about the expected source population at higher radio frequencies, such as those in the lower frequency bands of the \textit{Planck Surveyor} satellite. In particular, we have made empirical estimates of the source counts at 22~GHz, 30~GHz, 43~GHz and 70~GHz and compared these with both known data and current theoretical predictions. We have also made an estimate of the count at the ALMA frequency of 90~GHz, with a view to assessing the possible population of point sources available for the phase calibration of that instrument. 
\end{abstract}

\begin{keywords}
surveys -- cosmic microwave background -- radio continuum: general -- galaxies:evolution

\end{keywords}

\section{Introduction}
A major problem for many cosmic microwave background (CMB) experiments, such as the \textit{Planck Surveyor} mission, is the estimation of the contaminating effect of extragalactic radio sources on the measurement of the CMB anisotropies. In order to model the expected response from a CMB instrument, with a view to developing appropriate analysis techniques, it is essential to have a realistic assessment of the source count at the relevant frequency. There has been important work on modelling the source counts at high frequencies, such as that of Toffolatti et al. (1998) and De Zotti et al. (2005), but few direct observational measurements are available. In particular, in the bands of the \textit{Planck} Low Frequency Instrument (LFI) -- i.e. 30~GHz, 44~GHz, 70~GHz -- there is, at present, a lack of data from appropriate wide-field surveys, although several such surveys are under way: AT20G (Ricci et al. 2004, Sadler et al. 2006), OCRA (Browne et al. 2000), RATAN-600 (Parijskij 2004). The 9C survey (Waldram et al. 2003) at the comparatively close frequency of 15~GHz is therefore very important for making some predictions.

In this paper we have made use of data from the `simultaneous' multi-frequency 9C follow-up observations described in Bolton et al. 2004 (paper 1). There have already been two other papers based on these data: Bolton et al. 2006a (paper 2), which presents 5-GHz MERLIN and VLBA observations of a set of 36 compact sources, and Bolton et al. 2006b (paper 3), which reports results from a study of the 15-GHz variability of 51 sources. Here we have taken a complete sample of 110 sources above 25~mJy at 15~GHz and used the spectra over the range 1.4~GHz to 43~GHz to make some deductions about the expected source population in the \textit{Planck} LFI bands. Our approach is entirely empirical and there is no attempt to model the sources.

We have also used our data to make a prediction of the count at 90 GHz in order to estimate the number of point sources available as possible phase calibrators for the ALMA telescope. 

In section 2 we describe the sample and include some discussion of variability. Section 3 explains the principles of our method and the assumptions made. In section 4 we examine the spectral index distributions over the range 1.4 to 43~GHz and in section 5 we discuss the interpolation and extrapolation of the source spectra to 30, 70 and 90~GHz.  Section 6 presents the results of our calculations and section 7 attempts to estimate the errors involved. In section~8 we test our method by comparing our predictions for 1.4~GHz  and 4.8~GHz with the known counts at these frequencies. Section 9 presents estimates of the source counts at 22~GHz, 30~GHz, 43~GHz, 70~GHz and 90 GHz and compares them with those from the available high frequency data and with the theoretical models of De Zotti et al. (2005). Section 10 refers to the 90~GHz result and its implication for ALMA phase calibration. Finally, section 11 is a general discussion of the reliability of our method and section 12 is a summary of our conclusions.

We note here that our definition of spectral index, $\alpha$ for flux density, $S$, and frequency, $\nu$, is $S\propto \nu^{-\alpha}$. 

\section{The sample}

\begin{table}

\caption{The 110 sources in our sample with their simultaneous flux densities at 1.4, 4.8, 15.2, 22 and 43~GHz: $S_{1.4}$, $S_{4.8}$, $S_{15.2}$, $S_{22}$, $S_{43}$ in mJy}
\begin{tabular}{ccrrrrr}

\hline
    & Source name & $S_{1.4}$ & $S_{4.8}$ & $S_{15.2}$ & $S_{22}$ & $S_{43}$ \\

\hline

    1  &  J0002+2942 &   40.8 &   34.4 &     61.3 &   60.0 &   40.0 \\
    2  &  J0003+2740 &   54.1 &   70.0 &     67.2 &   56.2 &   19.0 \\
    3  &  J0003+3010 &   29.6 &   50.2 &     56.3 &   54.6 &   41.0 \\
    4  &  J0005+3139 &  755.5 &  268.1 &     81.6 &   58.0 &   25.0 \\
    5  &  J0010+3403 &  141.0 &   68.3 &     27.2 &   17.6 &    8.4 \\
    6  &  J0010+2838 &   64.9 &   48.5 &     46.4 &   55.0 &   49.7 \\
    7  &  J0010+2854 &   39.8 &   47.3 &     69.2 &  103.0 &  144.0 \\
    8  &  J0010+2717 &   63.3 &   34.8 &     31.8 &   32.0 &   17.7 \\
    9  &  J0010+2619 &  432.0 &  195.2 &     69.7 &   49.8 &   21.3 \\
   10  &  J0010+2956 &  209.8 &  114.0 &     58.9 &   49.8 &   26.0 \\
   11  &  J0010+2650 &   56.3 &   41.0 &     32.4 &   35.1 &   27.8 \\
   12  &  J0011+2803 &  583.4 &  186.3 &     49.1 &   32.0 &    6.4 \\
   13  &  J0011+2928 &  154.9 &   98.8 &     52.3 &   43.1 &   23.2 \\
   14  &  J0012+2702 &  638.0 &  219.0 &     73.9 &   51.0 &   14.0 \\
   15  &  J0012+3353 &   35.2 &   80.4 &    123.8 &  137.0 &  129.8 \\
   16  &  J0012+3053 &   18.5 &   22.1 &     25.5 &   27.7 &   20.1 \\
   17  &  J0013+2834 &   32.7 &   33.1 &     34.6 &   36.6 &   30.5 \\
   18  &  J0013+2646 &  364.0 &  118.6 &     30.0 &   17.0 &    5.0 \\
   19  &  J0014+2815 &   80.4 &   60.1 &     45.5 &   37.7 &   23.8 \\
   20  &  J0015+3216 & 1662.6 &  827.3 &    469.0 &  425.0 &  250.0 \\
   21  &  J0015+3052 &  225.1 &   90.0 &     38.0 &   24.5 &    8.8 \\
   22  &  J0018+2921 &  404.4 &  188.6 &     90.8 &   81.0 &   44.0 \\
   23  &  J0018+3105 &  364.0 &  119.0 &     45.0 &   17.0 &    6.0 \\
   24  &  J0018+2907 &   71.0 &   41.3 &     28.7 &   23.6 &   13.5 \\
   25  &  J0019+2817 &   25.9 &   23.8 &     17.4 &   27.0 &   34.9 \\
   26  &  J0019+2956 &   96.6 &   73.5 &     41.0 &   35.7 &   13.8 \\
   27  &  J0019+2647 &   98.5 &   70.1 &     66.7 &   73.6 &   63.1 \\
   28  &  J0019+3320 &   82.2 &   63.1 &     31.8 &   26.3 &   11.8 \\
   29  &  J0020+3152 &   25.7 &   42.7 &     31.8 &   19.1 &    8.0 \\
   30  &  J0021+2711 &  349.6 &  120.3 &     38.7 &   21.0 &   10.0 \\
   31  &  J0021+3226 &  179.8 &   82.4 &     27.4 &   15.8 &    4.0 \\
   32  &  J0022+3250 &   50.9 &   23.5 &     13.6 &   13.0 &    6.0 \\
   33  &  J0023+3114 &  140.9 &   66.8 &     32.1 &   24.6 &   12.0 \\
   34  &  J0023+2734 &  410.0 &  172.0 &     76.8 &   40.1 &   15.0 \\
   35  &  J0024+2911 &    5.0 &   20.5 &     42.3 &   34.0 &   15.0 \\
   36  &  J0027+2830 &  116.0 &   71.0 &     24.0 &   13.5 &    6.0 \\
   37  &  J0028+3103 &  165.0 &   66.0 &     36.1 &   15.0 &    8.0 \\
   38  &  J0028+2914 &  733.0 &  260.0 &     78.1 &   49.3 &    9.0 \\
   39  &  J0028+2954 &   22.0 &   24.0 &     24.9 &   18.6 &   15.0 \\
   40  &  J0029+3244 &  300.0 &  172.7 &     44.7 &   29.9 &   12.0 \\
   41  &  J0030+2957 &   24.0 &   14.5 &     13.5 &   16.0 &   12.2 \\
   42  &  J0030+3415 &   92.0 &   49.2 &     25.2 &   15.5 &    4.0 \\
   43  &  J0030+2833 &  572.0 &  170.7 &     47.4 &   24.0 &    6.0 \\
   44  &  J0031+3016 &  132.0 &   76.5 &     39.1 &   26.0 &    3.0 \\
   45  &  J0032+2758 &   28.1 &   34.1 &     30.1 &   21.6 &   12.0 \\
   46  &  J0033+2752 &  255.8 &   74.0 &     23.1 &    9.0 &    2.0 \\
   47  &  J0034+2754 &  820.0 &  471.4 &    295.0 &  236.7 &  129.2 \\
   48  &  J0036+2620 &  412.0 &  155.5 &     50.1 &   35.0 &   15.0 \\
   49  &  J0927+3034 &   52.0 &   44.0 &     47.0 &   37.0 &   18.6 \\
   50  &  J0928+2904 &  375.2 &  106.1 &     22.0 &   17.0 &    4.0 \\
   51  &  J0932+2837 &  102.5 &   95.8 &     58.4 &   46.8 &   27.0 \\
   52  &  J0933+2845 &  111.4 &   86.1 &     35.6 &   22.5 &   15.0 \\
   53  &  J0933+3254 &   46.0 &   36.6 &     22.8 &   21.9 &   14.0 \\
   54  &  J0936+3207 &   26.8 &   40.1 &     52.8 &   51.3 &   32.0 \\
   55  &  J0936+3313 &   56.9 &   48.3 &     29.6 &   30.3 &   19.0 \\
   56  &  J0937+3206 &  108.9 &   53.9 &     58.4 &   58.8 &   41.0 \\
   57  &  J1501+4211 &  121.5 &   54.0 &     28.3 &   20.0 &    8.0 \\
   58  &  J1502+3956 &  132.0 &   68.0 &     46.8 &   37.9 &   23.0 \\
   59  &  J1502+3947 &  403.9 &  130.0 &     35.2 &   21.3 &    3.0 \\
   60  &  J1502+3753 &  306.5 &  127.0 &     37.3 &   28.0 &   10.0 \\
   61  &  J1503+4528 &  498.4 &  159.0 &     65.9 &   41.0 &   14.3 \\
   62  &  J1505+3702 &  242.5 &   95.0 &     22.5 &   20.0 &    3.0 \\
   63  &  J1506+3730 & 1018.0 &  770.0 &    540.0 &  483.0 &  354.7 \\

\end{tabular}
\end{table}

\begin{table}
\contcaption{}
\begin{tabular}{ccrrrrr}

\hline
    & Source name & $S_{1.4}$ & $S_{4.8}$ & $S_{15.2}$ & $S_{22}$ & $S_{43}$ \\

\hline

   64  &  J1510+3750 &  731.0 &  300.0 &     76.9 &   43.0 &    8.8 \\
   65  &  J1510+4221 &  232.0 &  110.0 &     66.2 &   55.0 &   24.1 \\
   66  &  J1511+4430 &  344.0 &  110.0 &     62.5 &   48.0 &   22.0 \\
   67  &  J1514+3650 &  930.0 &  340.0 &     95.0 &   70.0 &   15.0 \\
   68  &  J1516+4349 &   28.3 &   25.1 &     21.8 &   18.5 &   15.2 \\
   69  &  J1516+3650 &  192.0 &  105.0 &     83.0 &   76.0 &   55.0 \\
   70  &  J1517+3936 &   16.6 &   26.0 &     40.3 &   43.0 &   35.3 \\
   71  &  J1518+4131 &   39.9 &   27.1 &     28.0 &   20.0 &   14.0 \\
   72  &  J1519+4254 &   69.9 &   67.2 &     99.9 &   97.0 &   82.0 \\
   73  &  J1519+3844 &   69.3 &   52.0 &     30.1 &   28.0 &   22.2 \\
   74  &  J1519+3913 &  241.0 &  103.0 &     37.6 &   26.0 &   10.0 \\
   75  &  J1520+3843 &  294.2 &  112.0 &     35.9 &   33.0 &   15.0 \\
   76  &  J1520+4211 &  124.1 &   56.5 &     53.8 &   71.8 &   85.0 \\
   77  &  J1521+4336 &  259.8 &  423.7 &    347.0 &  300.0 &  194.7 \\
   78  &  J1523+4156 &  555.5 &  137.0 &     56.9 &   41.0 &   12.0 \\
   79  &  J1525+4201 &  106.2 &   55.0 &     59.0 &   52.0 &   27.0 \\
   80  &  J1526+3712 &   47.9 &   72.0 &     64.6 &   64.0 &   37.8 \\
   81  &  J1526+4201 &   21.6 &   59.9 &     59.3 &   47.0 &   22.0 \\
   82  &  J1528+4219 &  216.5 &   84.0 &     43.6 &   37.0 &   15.0 \\
   83  &  J1528+4233 &  138.6 &   57.2 &     32.2 &   24.0 &    8.0 \\
   84  &  J1528+3738 & 1051.8 &  336.0 &     74.2 &   53.0 &   18.8 \\
   85  &  J1528+3816 &   22.6 &   46.0 &     72.5 &   86.0 &   71.2 \\
   86  &  J1528+4522 &  173.9 &   75.0 &     43.6 &   47.0 &   25.0 \\
   87  &  J1529+4538 &  278.0 &  110.0 &     36.3 &   23.2 &    8.0 \\
   88  &  J1529+3945 &  134.6 &   60.0 &     29.8 &   28.0 &   24.5 \\
   89  &  J1530+3758 &  103.1 &  135.0 &     60.8 &   38.0 &   13.1 \\
   90  &  J1531+4356 &   53.1 &   55.0 &     25.8 &   18.0 &   17.0 \\
   91  &  J1531+4048 &  338.3 &  105.0 &     31.2 &   26.0 &    6.0 \\
   92  &  J1533+4107 &   19.1 &   17.5 &     19.9 &   19.5 &   13.3 \\
   93  &  J1538+4225 &   42.0 &   40.5 &     41.6 &   42.2 &   29.0 \\
   94  &  J1539+4217 &   53.3 &   40.0 &     34.2 &   37.0 &   26.3 \\
   95  &  J1540+4138 &   16.0 &   30.9 &     34.2 &   23.0 &    9.1 \\
   96  &  J1541+4114 &   65.0 &   38.0 &     30.5 &   27.5 &   19.2 \\
   97  &  J1541+4456 &  377.8 &  131.0 &     47.6 &   30.0 &    7.8 \\
   98  &  J1545+4130 &   72.0 &   56.8 &     50.0 &   45.0 &   22.2 \\
   99  &  J1546+4257 &  347.0 &  110.0 &     33.2 &   27.0 &   15.0 \\
  100  &  J1547+4208 &   72.7 &   71.0 &     56.5 &   47.0 &   17.8 \\
  101  &  J1548+4031 &   61.6 &   60.0 &     84.9 &   72.3 &   37.0 \\
  102  &  J1550+4536 &   47.6 &   60.0 &     31.8 &   20.0 &    3.0 \\
  103  &  J1550+4545 &   23.0 &   17.9 &     17.9 &   18.0 &   12.1 \\
  104  &  J1553+4039 &   47.5 &   43.0 &     33.3 &   18.6 &    8.8 \\
  105  &  J1554+4350 &    6.5 &   33.2 &     41.5 &   37.0 &   20.5 \\
  106  &  J1554+4348 &   52.8 &   60.8 &     44.8 &   38.0 &   15.7 \\
  107  &  J1556+4259 &   63.0 &   94.0 &     57.8 &   43.0 &   20.0 \\
  108  &  J1557+4522 &  509.2 &  262.0 &    111.3 &   85.0 &   40.9 \\
  109  &  J1557+4007 &  101.4 &   87.0 &     79.4 &   74.0 &   39.6 \\
  110  &  J1558+4146 &  238.0 &   75.0 &     32.8 &   17.7 &    7.0 \\

\hline

\end{tabular}
\end{table} 

\begin{figure}
        {\epsfig{file=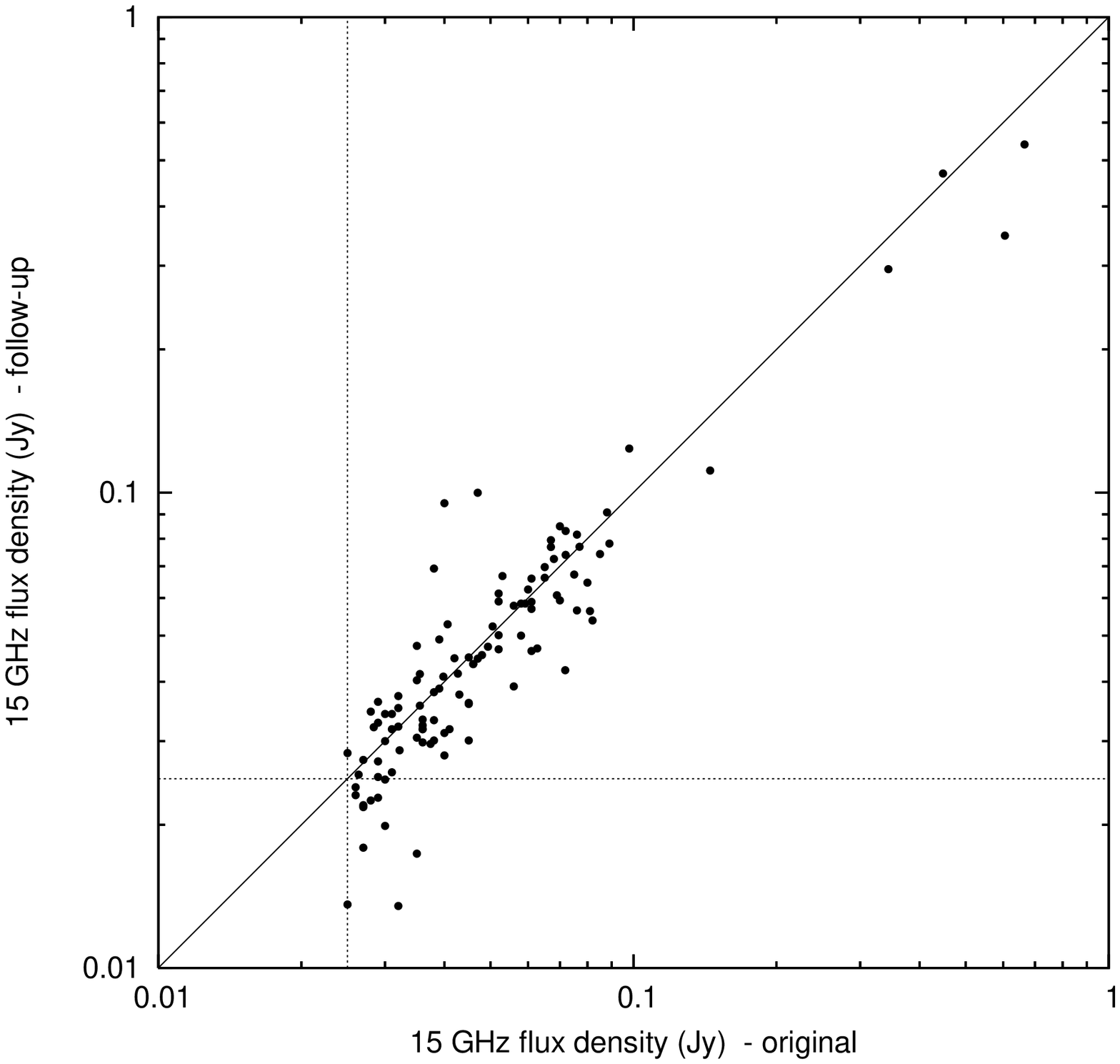,
        angle=270,
        width=7.5cm,clip=}} 
        \caption{Plot of 15~GHz follow-up flux densities versus original flux densities, showing the scatter about the line of unit slope. The dotted lines correspond to a flux density of 25 mJy on each axis.}
\end{figure}

We assembled a flux-limited sample of 121 sources, complete to 25~mJy at 15~GHz, from three areas of the 9C survey. All the sources had a complete set of simultaneously measured flux densities at frequencies of 1.4~GHz, 4.8~GHz, 15.2~GHz, 22~GHz and 43~GHz apart from 11 of them, for which some observations were missing. Since the reasons for the omissions were totally unrelated to the source characteristics -- i.e. they were due to weather or scheduling problems -- we have simply omitted these sources from the sample and calculated the final effective area from a fit of the known 15~GHz source count (see below). We thus have a `snapshot' of sets of instantaneous spectra for a complete sample of 110 sources, in an area of approximately 130 deg$^{2}$, as listed in Table 1. (For more details of individual sources see papers 1 and 2.)

The extent of variability at 15 GHz, as illustrated by our sample, can be seen in Figure 1 which shows how the sources have changed in flux density beween the time of the original observations and the time of the follow-up observations. The sources in the original selection had observation dates ranging from November 1999 to September 2001 and the follow-up observations had dates in January, November, December 2001 and January, May 2002. The maximum time interval between the original and follow-up observations of any source was 30 months and the minimum was 2 months. The source showing maximum variability was J1514+3650 which had risen from 40 mJy to 95 mJy in a period of 19 months. Since the time intervals between our original and follow-up observations differ widely, Figure 1 can give only a general indication of the extent of variability at 15 GHz. (For more detailed work see paper 3.)

Variability presents a problem for the selection of a flux limited sample, since close to the lower limit there will be preferential selection of those sources which were above their mean flux density values at the time of observation rather than below. Hence, for our sample, Figure 1 shows an excess of sources with flux densities below 25 mJy in the follow-up observations, and this is, in fact, unavoidable.

Another feature of our sample is that it contains significantly fewer sources above 100~mJy than are expected from the known 15 GHz source count (Waldram et al. 2003). This became apparent when we were calculating the effective sample area from a fit of the known count to the flux density distribution in the original sample. We found that in the range from 100~mJy to the maximum value of 665~mJy there were only 5 sources and we therefore used only those below 100~mJy in our fit. (In fact, a value for this area is not required for our predictions and is used only in Section 7 and Figure 7  for calculating the incomplete counts derived directly from the sample, by way of comparison.) 

Similarly, in the follow-up sample there are only 6 sources above 100~mJy, the highest value being 540~mJy. Taking account of the uncertainty in the area and the Poisson errors, we estimate that we should expect at least twice this number. The short-fall may be be due to the fact that, in the original selection of the 9C fields for the purpose of CMB observations, there was some bias against regions predicted to contain very bright sources (see Waldram et al. 2003). However, our area here is quite small and, ideally, a much larger area is required to sample adequately the population above 100~mJy.

Thus, although for the purpose of this work we can assume that our original sample represents the typical characteristics of a complete sample of sources in the flux density range 25~mJy to 100~mJy, there is less certainty in the under-sampled upper range, above 100~mJy. The implications of this are discussed in later sections of the paper.

\section{Empirical estimation of the source counts}
We now consider whether, knowing the source count at 15 GHz, it is possible to estimate the source counts at higher frequencies from these data. It is not possible to do so directly because a complete sample would be far too small. For example, if we assume an extreme rising spectral index between 15 and 43~GHz, $\alpha_{15}^{43}$, of $-1$, our sample at 43 GHz is complete to only $\sim 70$ mJy, providing only 9 sources. It is however possible to use the 15 GHz count and our $\alpha_{15}^{43}$ distribution to estimate the count at 43 GHz, if we make certain assumptions about the $\alpha_{15}^{43}$ distribution in our flux-density range.

Consider first a source population such that each source has the same spectral index $\alpha$ between the two frequencies $\nu_{1}$ and $\nu_{2}$, or $S_{\nu_{1}} = r S_{\nu_{2}}$ where $r = \left({\nu_{1}}/{\nu_{2}}\right)^{-\alpha}$, and 
let us assume that the differential count at $\nu_{1}$ has the form
 \[n_{\nu_{1}}(S) = {\rm d}N_{\nu_{1}}/{\rm d}S = A_{\nu_{1}} S ^{\rm -b}\]
where  $A_{\nu_{1}}$ and ${\rm b}$ are constants.\\
In order to find the differential count at $\nu_{2}$ we consider the corresponding integrated counts, $N_{\nu_{1}}(>S)$ and $N_{\nu_{2}}(>S)$. At frequency~$\nu_{1}$ (for $\rm b \neq 1$)
 \[N_{\nu_{1}}(>S) = A_{\nu_{1}}(1-{\rm b})^{-1}S^{\rm 1-b}\]
and so at frequency $\nu_{2}$
 \[N_{\nu_{2}}(>S) = N_{\nu_{1}}(>rS) = A_{\nu_{1}}(1-{\rm b})^{-1}(rS)^{\rm 1-b}.\]
This means that the differential count at $\nu_{2}$ becomes
 \[n_{\nu_{2}}(S) = {\rm d}N_{\nu_{2}}/{\rm d}S = r^{1-{\rm b}}A_{\nu_{1}}S^{-{\rm b}}.\] 

In practice, however, we know that the source population spans a range of spectral indices and we now make the following assumptions: first, that our sample provides a typical distribution of spectral indices and secondly, that this distribution is independent of flux density.
In our sample of $m$ sources we know the spectral index $\alpha_{i}$ and the corresponding value of $r$, $r_{i}$, for each source, and so can calculate  $k_{i} = r_{i}^{1-{\rm b}}$, and hence
 \[n_{\nu_{2}}(S) = K A_{\nu_{1}} S ^{\rm -b}\]
where
 \[K = \frac{1}{m} \sum_{i=1}^{m} k_{i} = \frac{1}{m} \sum_{i=1}^{m} r_{i}^{1-{\rm b}}\]
(See Condon 1984 and Kellermann 1964 for similar analyses.)

We see that these assumptions lead to a form of the count at $\nu_{2}$ with the same exponent as the count at $\nu_{1}$ but with a different prefactor. 
We can define an effective value for $r$, $r_{e}$, such that
 \[K = r_{\rm e}^{1-{\rm b}}\]
where $r_{e}$ would be the value of $r$ for all sources, if they all had the same spectral index. Using $r_{e}$, we can then estimate the flux density range over which the count derived for frequency $\nu_{2}$, from a knowledge of the count at frequency $\nu_{1}$ (i.e. at 15~GHz), may be assumed to be valid. For this purpose we define values at $\nu_{2}$: $S_{\rm min} = {S^{15}_{\rm min}}/r_{\rm e}$, $S_{\rm max} = {S^{15}_{\rm max}}/r_{\rm e}$ and 
$S_{\rm c} = {S^{15}_{\rm c}}/r_{\rm e}$, where ${S^{15}_{\rm min}}$  and ${S^{15}_{\rm max}}$ are the minimum and maximum values in the original 15~GHz sample, 25mJy and 665 mJy respectively, and ${S^{15}_{\rm c}}$ is the upper `completeness' value of 100~mJy above which we know the data are sparse. We assume that, although our predictions may be reliable in the range $S_{\rm min}$ to $S_{\rm c}$, they will be less so in the range $S_{\rm c}$~to $S_{\rm max}$.

\section{Spectral Indices in range 1.4 to 43 GHz}

\begin{figure*}
        {\epsfig{file=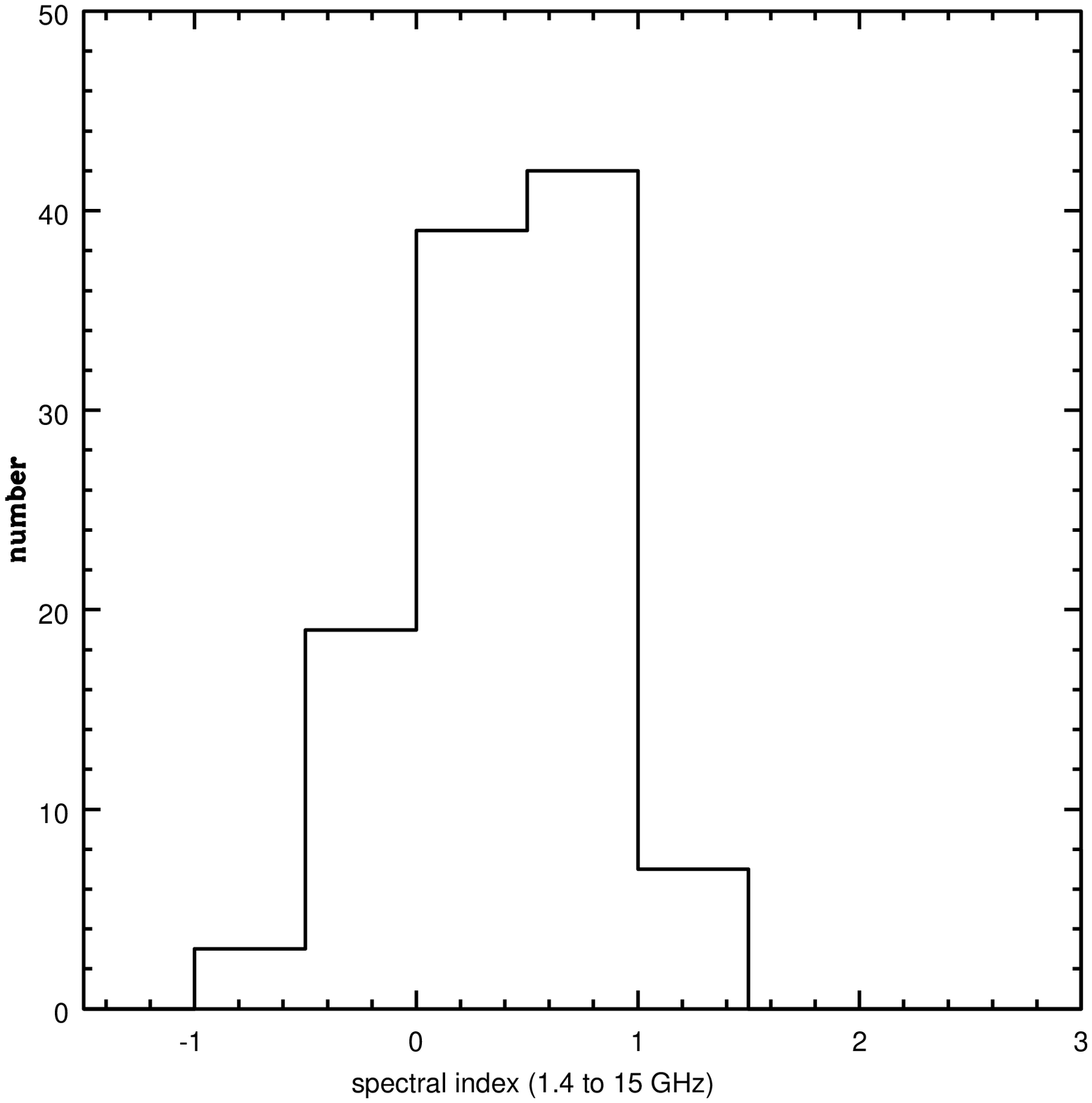,
        width=7.0cm,clip=}} 
        {\epsfig{file=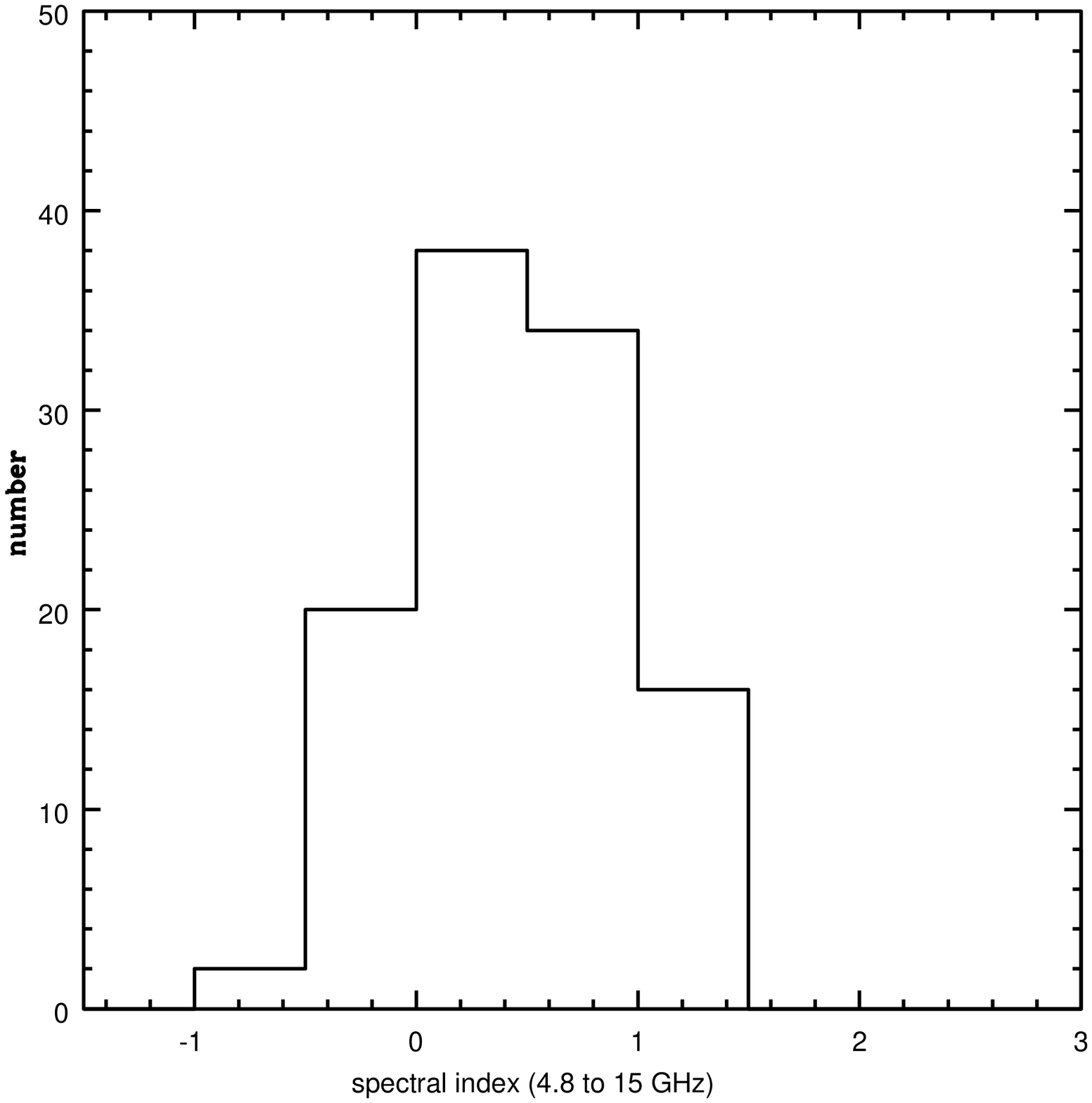,
        width=7.0cm,clip=}} 
        {\epsfig{file=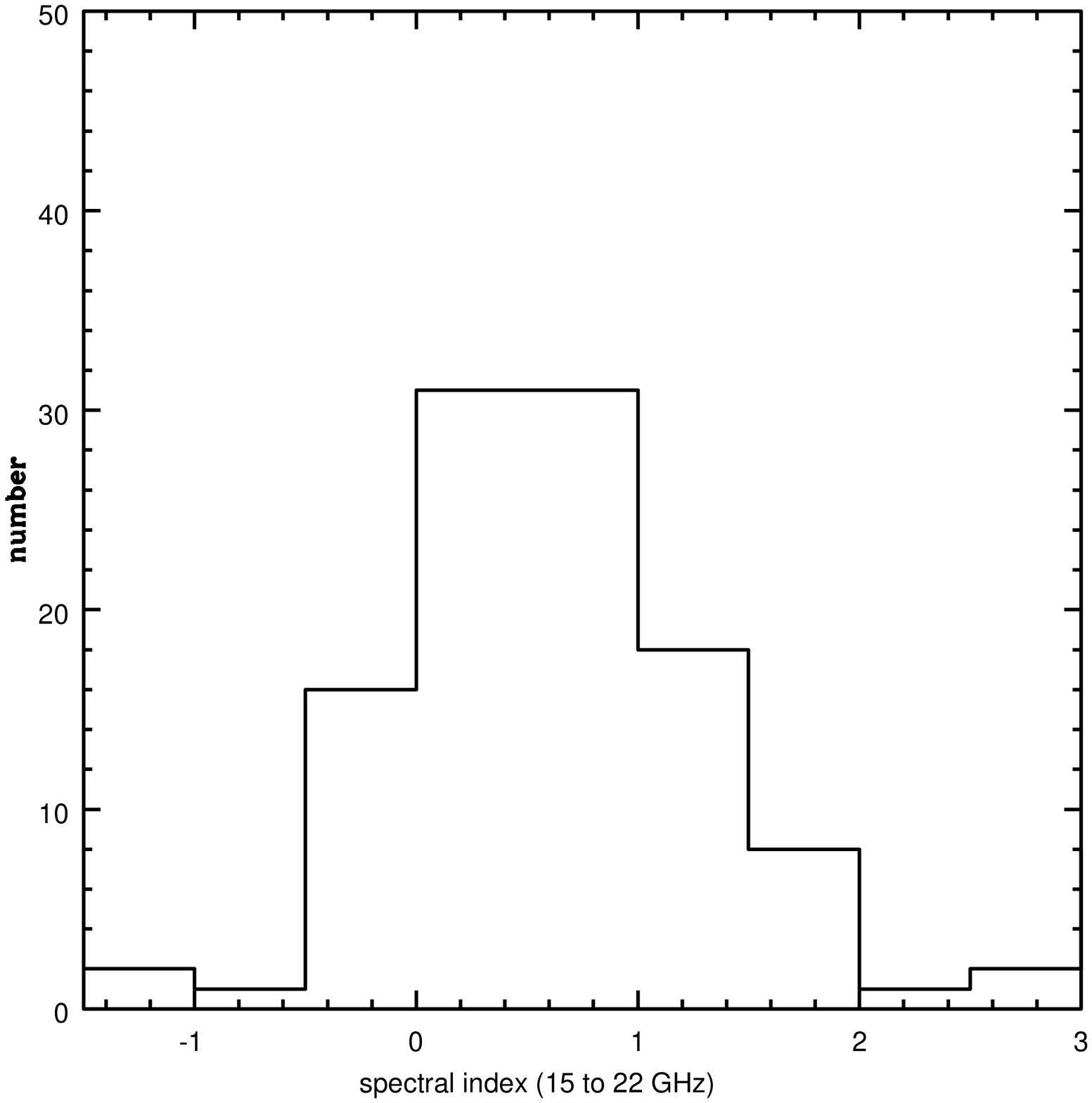,
        width=7.0cm,clip=}} 
        {\epsfig{file=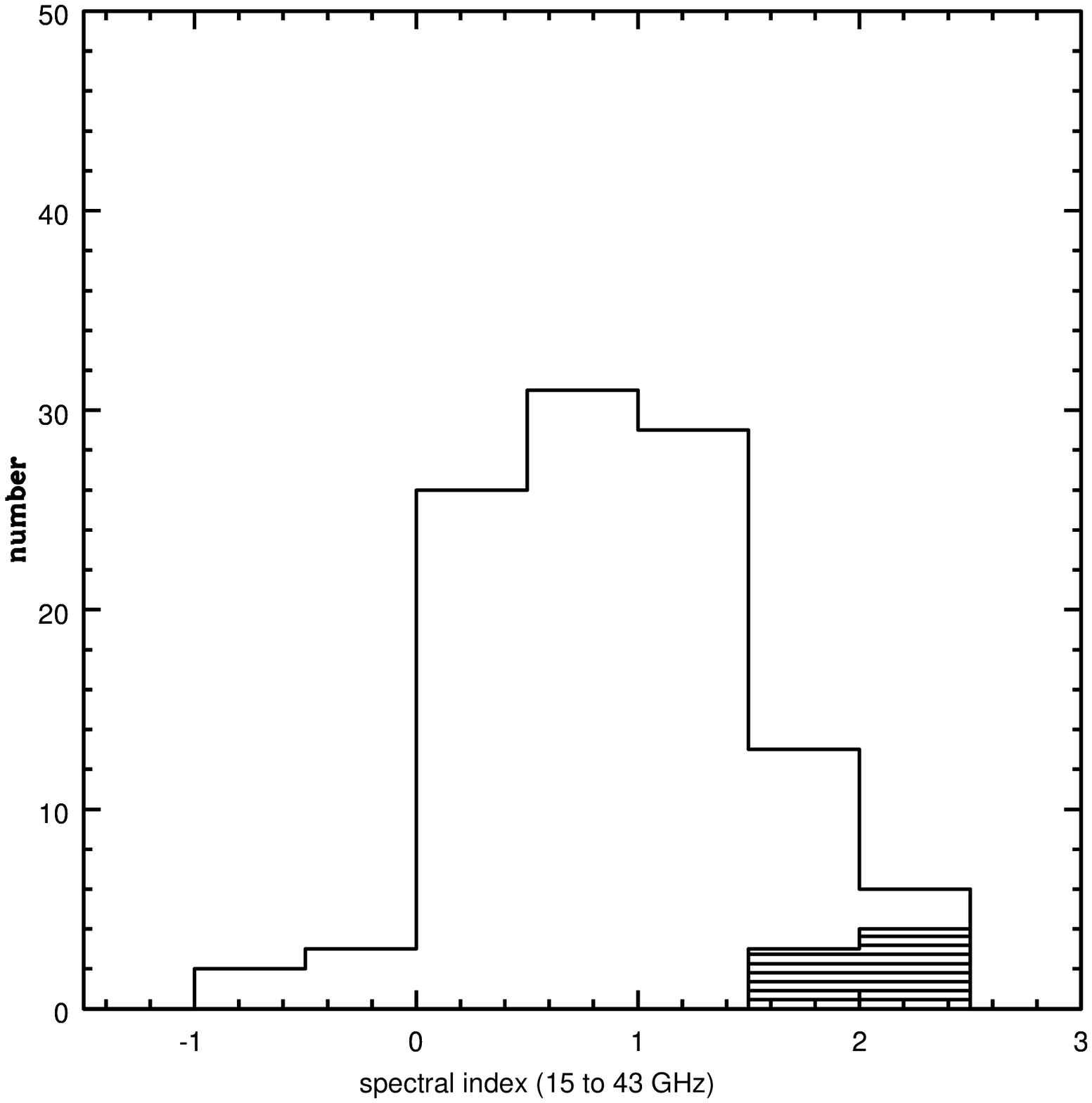,
        width=7.0cm,clip=}} 
        \caption{Distributions of spectral index: $\alpha_{1.4}^{15}$, $\alpha_{4.8}^{15}$, $\alpha_{15}^{22}$, $\alpha_{15}^{43}$.  The shaded area indicates the 7 non-detections at 43~GHz where the flux densities have been set equal to the noise level.}
\end{figure*}

\begin{figure*}
        {\epsfig{file=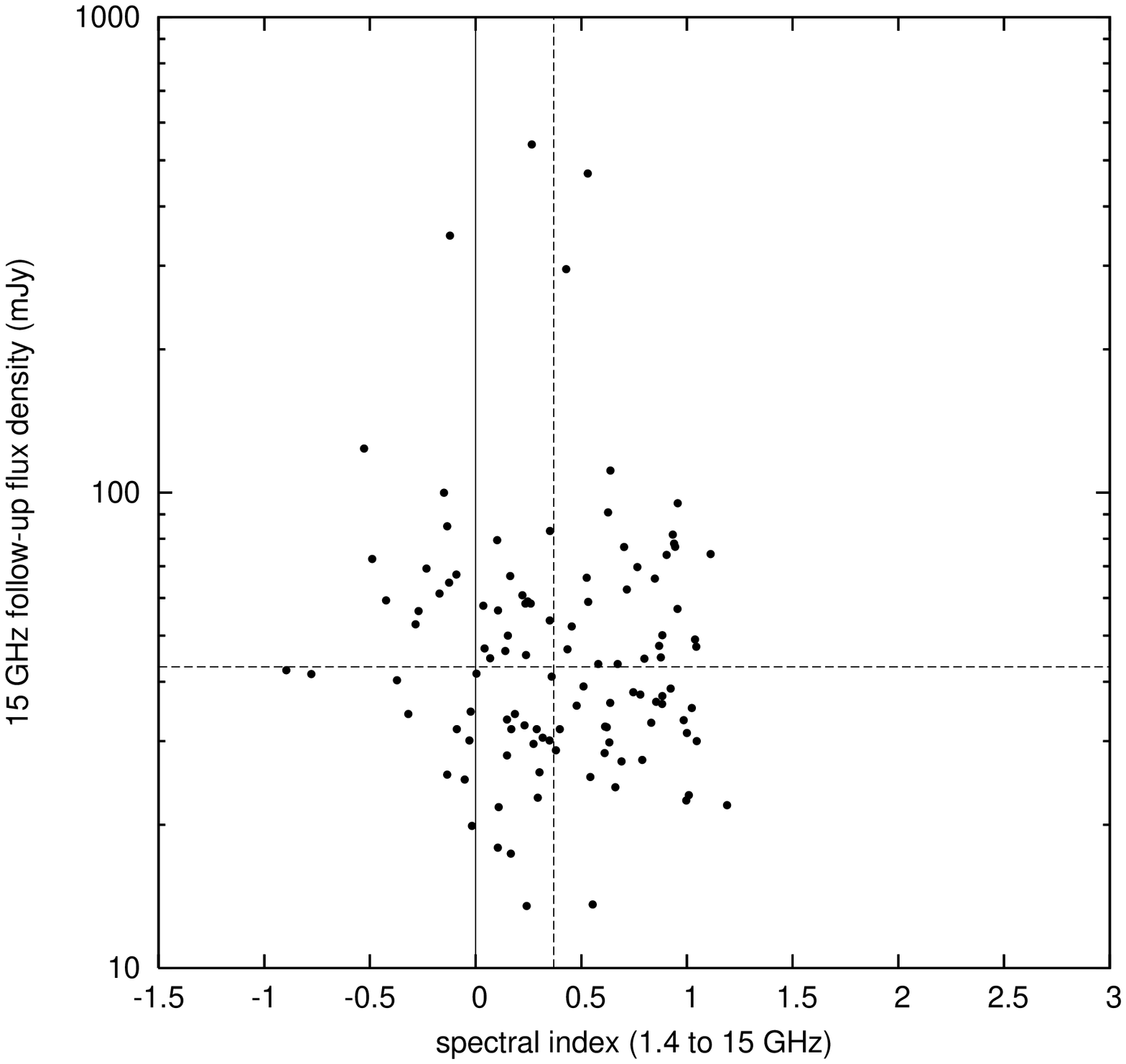,
        angle=270,width=7.0cm,clip=}} 
        {\epsfig{file=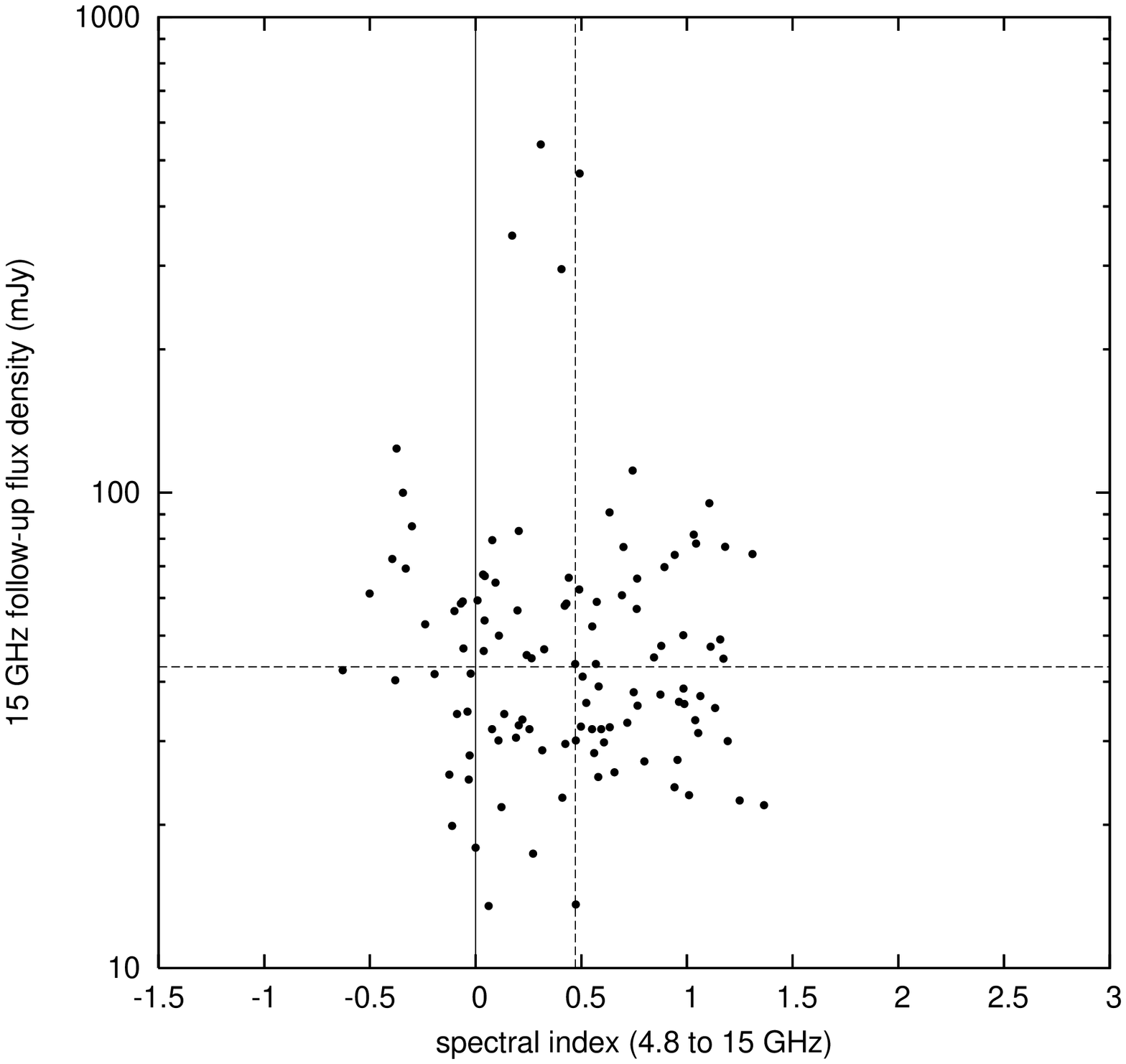,
        angle=270,width=7.0cm,clip=}} 
        {\epsfig{file=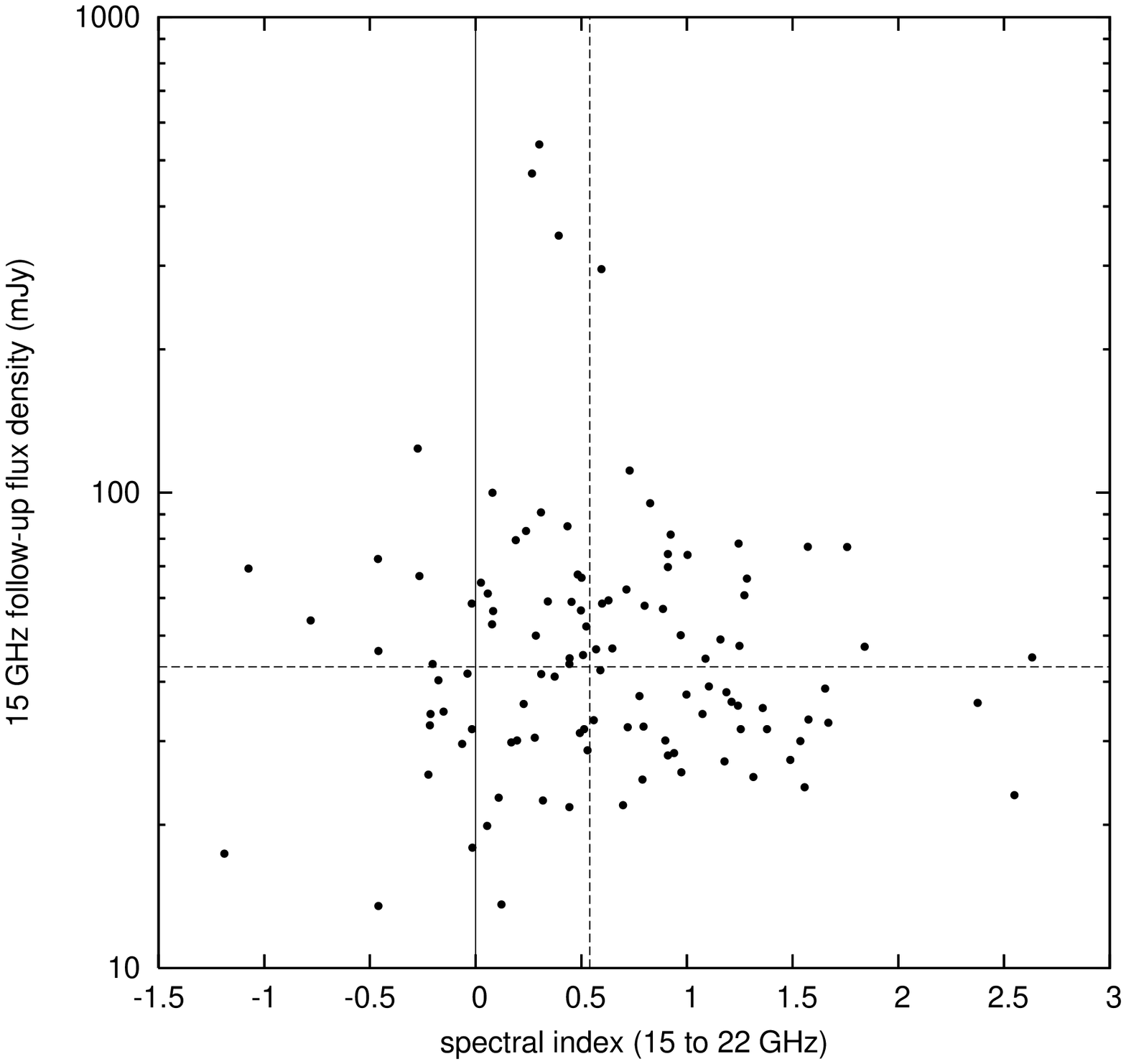,
        angle=270,width=7.0cm,clip=}} 
        {\epsfig{file=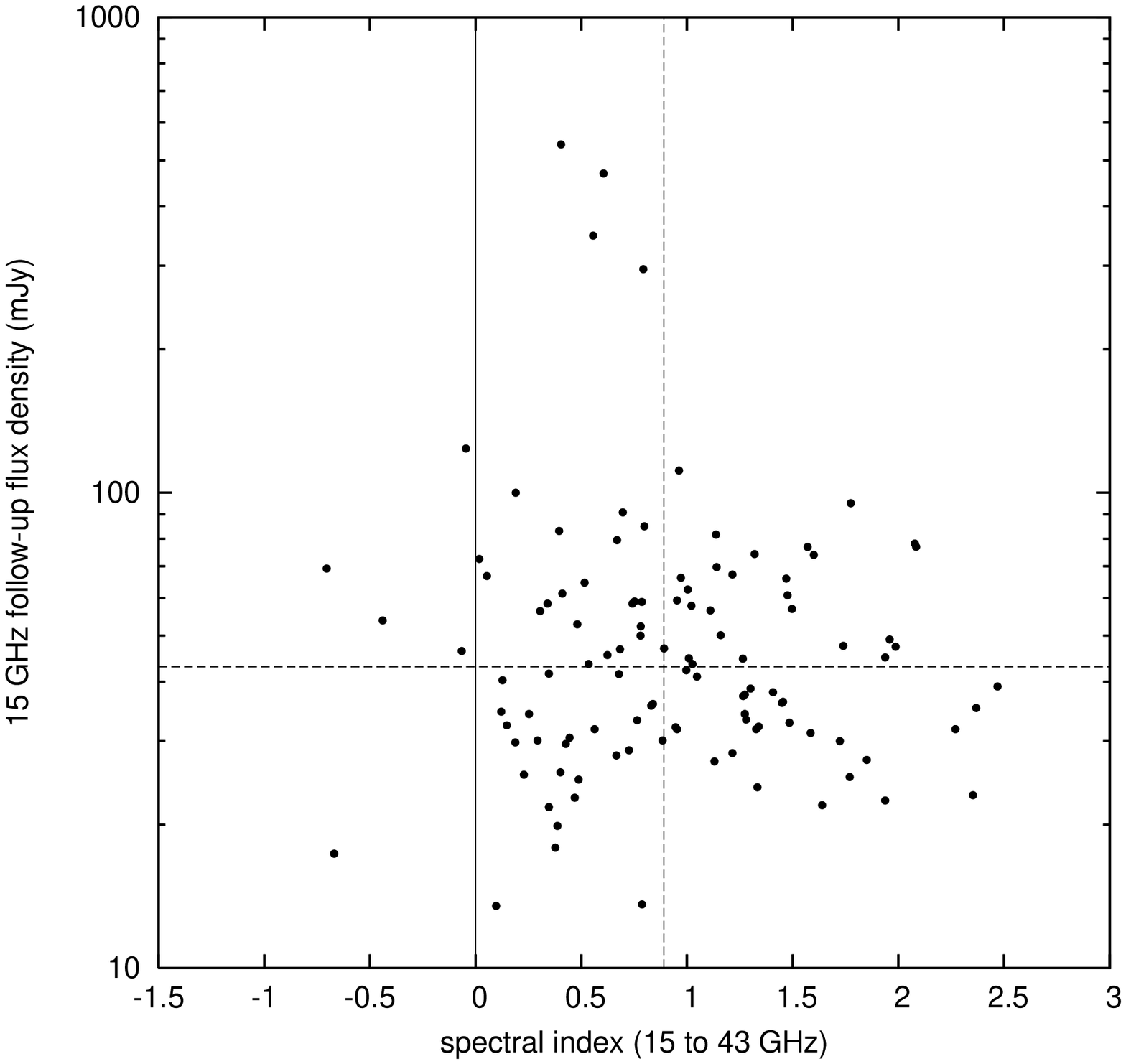,
        angle=270,width=7.0cm,clip=}} 
        \caption{Scatter plots of the 15~GHz follow-up flux density versus spectral index for: $\alpha_{1.4}^{15}$, $\alpha_{4.8}^{15}$, $\alpha_{15}^{22}$, $\alpha_{15}^{43}$. In each case the dashed lines show the median flux density and median spectral index. See Table 2.}
\end{figure*}

\begin{figure}
        {\epsfig{file=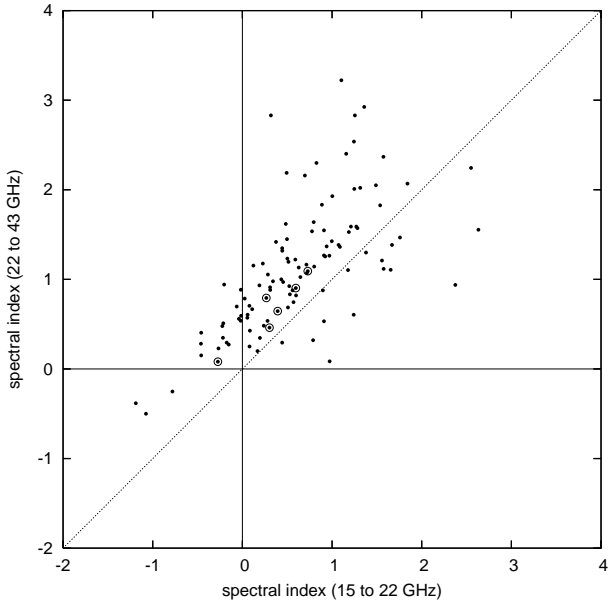,
        angle=270,width=8.0cm,clip=}} 
        \caption{Plot of $\alpha_{22}^{43}$ versus $\alpha_{15}^{22}$, showing the line corresponding to $\alpha_{22}^{43}~=~\alpha_{15}^{22}$. Sources with 15 GHz (follow-up) flux densities above 100~mJy are ringed.}
\end{figure}

\begin{table}
\caption{Table showing the distribution of numbers of sources with respect to the median spectral index ($\rm{\alpha_{med}}$) and the median 15 GHz follow-up flux density ($\rm{S_{med}} = 43.0 mJy $). See Figure 2.}
\begin{tabular}{cccccc}
\hline
Frequency & $\rm{\alpha_{med}}$ & $S>\rm{S_{med}}$ & $S>\rm{S_{med}}$ & $S<\rm{S_{med}}$ & $S<\rm{S_{med}}$ \\
GHz &  & $\alpha<\rm{\alpha_{med}}$ & $\alpha>\rm{\alpha_{med}}$ &  $\alpha<\rm{\alpha_{med}}$ & $\alpha>\rm{\alpha_{med}}$ \\
\hline

     1.4  &   0.37   &  28  &  27  &  27  &  28 \\
     4.8  &   0.472  &  31  &  24  &  24  &  31 \\
    22.0  &   0.54   &  30  &  25  &  25  &  30 \\
    43.0  &   0.89   &  28  &  27  &  27  &  28 \\

\hline
\end{tabular}
\end{table}

Figure 2 shows histograms of the spectral index distributions in the range 1.4 to 43~GHz. In the case of the 43 GHz observations there were 7 non-detections of sources. For these we have set the flux density value equal to the noise level and the corresponding spectral indices are shown shaded in the figure. It can be seen that the median spectral index increases with increasing frequency, being 0.37 for $\alpha_{1.4}^{15}$, 0.47 for $\alpha_{4.8}^{15}$, 0.54 for $\alpha_{15}^{22}$ and 0.89 for $\alpha_{15}^{43}$. In Figure 3, in order to investigate the dependence of spectral index on flux density, we have drawn scatter plots of the 15 GHz follow-up flux density versus spectral index for the four frequencies and added lines showing the median values of the two quantities. There is no apparent correlation visible and this is confirmed by Table~2 which gives the number of sources in each of the four quadrants of each plot. We conclude that, considering the Poisson errors in these small numbers, the differences are not significant, although, as we have seen in Section 2, the distributions are undersampled above 100 mJy. 

In Figure~4 we have drawn a scatter plot of $\alpha_{22}^{43}$ versus $\alpha_{15}^{22}$. Sources with 15 GHz (follow-up) flux densities above 100~mJy are ringed. It can be seen that only three sources have rising spectra from 22 to 43~GHz: these are J0010+2854, J0019+2817 and J1520+4211 and have 15 GHz (follow-up) flux densities of 69.2, 17.4 and 53.8~mJy respectively, all, in fact, less than 100~mJy.

For further discussion of the reliability of the spectral index measurements see section 11.

\section{Interpolation and extrapolation of the source spectra}

\begin{figure}
        {\epsfig{file=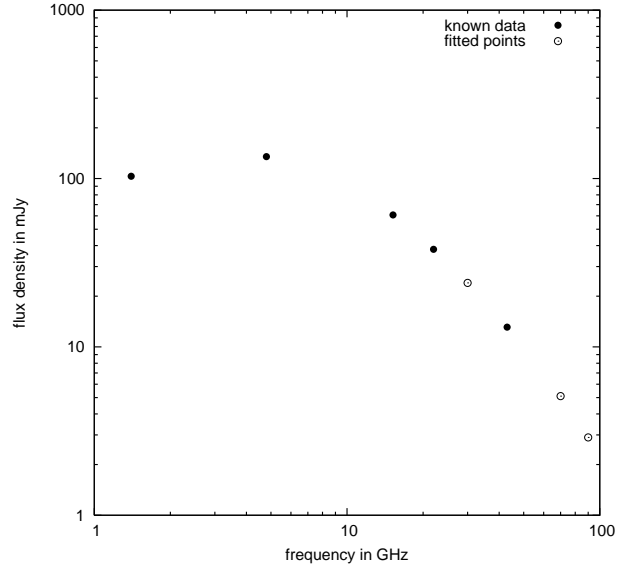,
        angle=270,width=8.0cm,clip=}} 
        \caption{Spectrum of J1530+3758, showing the known data points and the points calculated from a quadratic fit of $\log{S}$ with $\log{\nu}$.}
\end{figure}

In order to estimate the value of $K$ for the frequencies 30, 70 and 90~GHz, we
have needed to interpolate and extrapolate the spectra of the individual sources by an appropriate fit to the existing data. From the `colour-colour' plot in Figure~4 we see that, out of a total of 110 sources, 91 have both $\alpha_{15}^{22}$ and $\alpha_{22}^{43}$ $>0$ and we assume that these are steep spectrum sources whose spectra have already turned over at frequencies below 43 GHz. Of these, 75 have $\alpha_{22}^{43}$ $>$ $\alpha_{15}^{22}$ and so any fit should attempt to take account of this steepening of their spectra (see, for example, the plot for J1530+3758 in figure 5).

We have tried three types of fit of $\log{S}$ with $\log{\nu}$ : (a) a quadratic fit for all sources using the values at 15.2, 22 and 43~GHz , (b) a similar quadratic fit for the 75 sources mentioned above but a linear fit for the remainder, using the 22 and 43~GHz values and (c) a linear fit for all sources, using the 22 and 43~GHz values. The linear fit is equivalent to the assumption that $\alpha_{43}^{70}$ and $\alpha_{43}^{90}$ are both equal to $\alpha_{22}^{43}$.

In each case, our procedure was to apply the appropriate fit to the known spectral points of each source and calculate the flux densities at 30~GHz, 70~GHz and 90~GHz by interpolation or extrapolation. We could then find the values of $r_{i}$ at each frequency and hence the corresponding values of $K$.

We examined the results of the three types of fit for a selection of sources from the different areas in the `colour-colour' plot. Fit (a) catered well for the 75 sources with steepening falling spectra but produced some obvious anomalies for a number of the flatter spectra. Fit (b) appeared to produce much more appropriate results for all types of spectra, while fit (c) did not take sufficient account of the steepening falling spectra. It was therefore decided to use the values of $K$ from fit (b), but the results from all three types of fit are included in Table~5.

We should emphasise that this method of interpolation and extrapolation is purely empirical and no attempt has been made to model the sources or to examine every source individually.

\section{The calculations}

\begin{table}
\caption{Table showing the values of $K$ and their uncertainties for the four frequencies 1.4, 4.8, 22 and 43~GHz.}

\begin{tabular}{ccccccc}
\hline
Frequency & $K$ & $K$ & $K$  \\
GHz & all sources & $S_{15}<43$ mJy & $S_{15}>43$ mJy \\
\hline

     1.4 & $5.38\pm0.53$ & $5.51\pm0.77$ & $5.25\pm0.73$ \\ 
     4.8 & $2.15\pm0.13$ & $2.23\pm0.18$ & $2.08\pm0.18$ \\
     22 & $0.80\pm0.02$ & $0.78\pm0.03$ & $0.82\pm0.03$ \\
     43 & $0.44\pm0.04$ & $0.42\pm0.05$ & $0.45\pm0.05$ \\

\hline
\end{tabular}

\end{table}

\begin{table}
\caption{Table showing the values of $K$ and their uncertainties for the three frequencies 30, 70 and 90~GHz. See section 5.}

\begin{tabular}{ccccccc}
\hline
& & $K$ & $K$ & $K$  \\
& & all sources & $S_{15}<43$ mJy & $S_{15}>43$ mJy \\
\hline

  30 GHz &  &  &  & \\ 
  &  (a) & $0.62\pm0.03$ & $0.60\pm0.04$ & $0.65\pm0.05$ \\
  &  (b) & $0.61\pm0.03$ & $0.59\pm0.04$ & $0.63\pm0.04$ \\
  &  (c) & $0.59\pm0.03$ & $0.57\pm0.04$ & $0.61\pm0.04$ \\
  70 GHz &  &  &  & \\
  &  (a) & $0.25\pm0.03$ & $0.25\pm0.04$ & $0.25\pm0.05$ \\
  &  (b) & $0.29\pm0.04$ & $0.29\pm0.06$ & $0.30\pm0.07$ \\
  &  (c) & $0.33\pm0.04$ & $0.31\pm0.05$ & $0.34\pm0.07$ \\
  90 GHz &  &  &  & \\
  &  (a) & $0.19\pm0.03$ & $0.21\pm0.04$ & $0.18\pm0.04$ \\
  &  (b) & $0.25\pm0.05$ & $0.25\pm0.06$ & $0.26\pm0.08$ \\
  &  (c) & $0.29\pm0.05$ & $0.27\pm0.06$ & $0.31\pm0.08$ \\

\hline
\end{tabular}

\end{table}

\begin{table}
\caption{Table showing the results of our calculations for the frequencies from 1.4~GHz to 90~GHz}
\begin{tabular}{ccccrrr}
\hline
Frequency & $K$ & $A$ & $r_{\rm e}$ & $S_{\rm min}$ & $S_{\rm max}$ & $S_{\rm c}$  \\
GHz & & $\rm{Jy^{-1}sr^{-1}}$ & & mJy & mJy & mJy \\
\hline

     1.4 & $5.38$ & $274\pm30$ & 0.23  &  110 & 2900 & 435 \\
     4.8 & $2.15$ & $110\pm9$  & 0.51  &   49 & 1300 & 195 \\
    15.2 & $1.00$ &  $51\pm3$  & 1.00  &   25 &  665 & 100 \\
    22.0 & $0.80$ &  $41\pm2$  & 1.21  &   21 &  550 &  85 \\
    30.0 & $0.61$ &  $31\pm2$  & 1.54  &   16 &  430 &  65 \\
    43.0 & $0.44$ &  $22\pm2$  & 2.04  &   12 &  330 &  50 \\
    70.0 & $0.29$ &  $15\pm2$  & 2.93  &    9 &  230 &  35 \\
    90.0 & $0.25$ &  $13\pm3$  & 3.34  &    7 &  200 &  30 \\

\hline
\end{tabular}
\end{table}

\begin{figure}
        {\epsfig{file=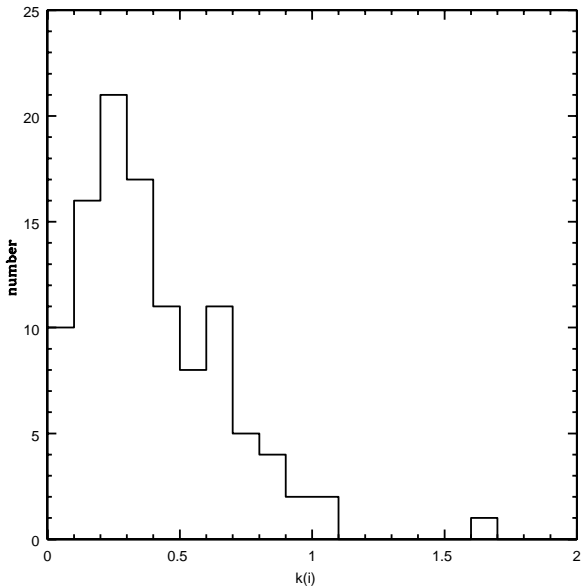, width=8.0cm, clip=}}
        \caption{Distribution of $k_{i}$ values for 43 GHz.}
\end{figure}

For each frequency we have calculated the value of $k_{i}$ for each source, where $k_{i} = r_{i}^{\rm 1-b}$, and then taken the unweighted mean of the distribution,  
$K =  \frac{1}{m} \sum_{i=1}^{m} k_{i}$. This was repeated with the sample divided into two groups, one with $S_{15} < 43$ mJy and the other with $S_{15} > 43$ mJy, 43~mJy being the median flux density of the 15~GHz follow-up observations. The results for the frequencies 1.4, 4.8, 22 and 43~GHz are shown in Table 3.  In Table 4, for 30, 70 and 90~GHz, we have also included the results from the three methods of spectral interpolation/extrapolation, (a), (b), (c), described above. In each case the uncertainty quoted is the error in the mean, or $\sigma/\sqrt{m}$ , where $\sigma$ is the standard deviation of the $k_{i}$ distribution and $m$ is the total number of sources. The distributions are necessarily skewed; an example is shown in Figure 6, for 43 GHz. 

Table~5 shows the results of our calculations. We take the count at 15 GHz to be
\[n(S) \equiv \frac{{\rm d}N}{{\rm d}S} \approx 51 \left( \frac{S}{\rm Jy} \right)^{-2.15}
\, {\rm Jy}^{-1}{\rm sr}^{-1}
\]
(Waldram et al. 2003). At another frequency $\nu$ the exponent $\rm{b}$~(~$=~2.15$~) remains the same but the prefactor $A$ becomes $K\times51$. For each frequency we have calculated $r_{\rm e}$, and also the values of $S_{\rm min}$, $S_{\rm max}$ and $S_{\rm c}$.

\begin{figure*}
        {\epsfig{file=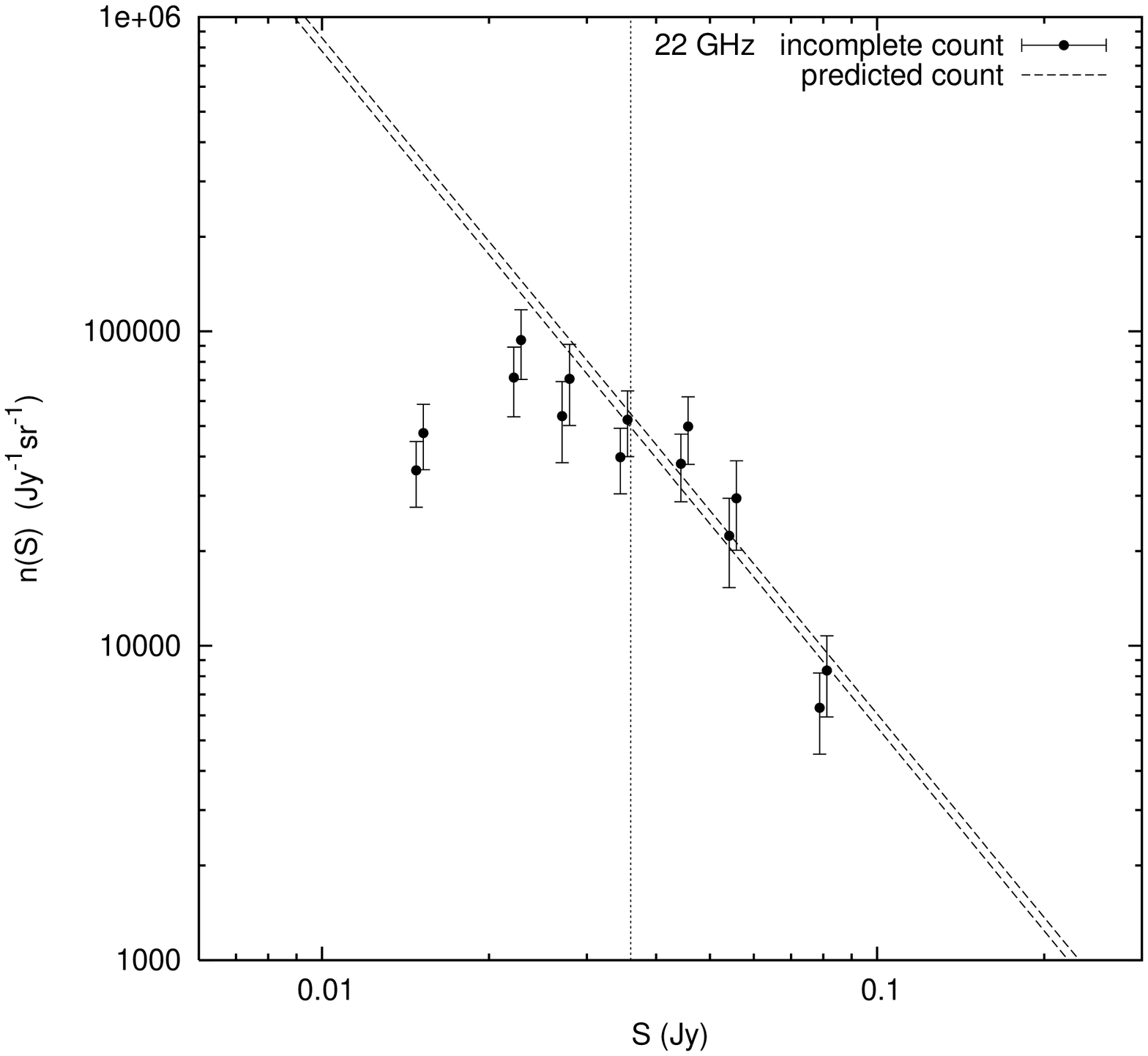,
        angle=270,width=8.0cm,clip=}} 
        {\epsfig{file=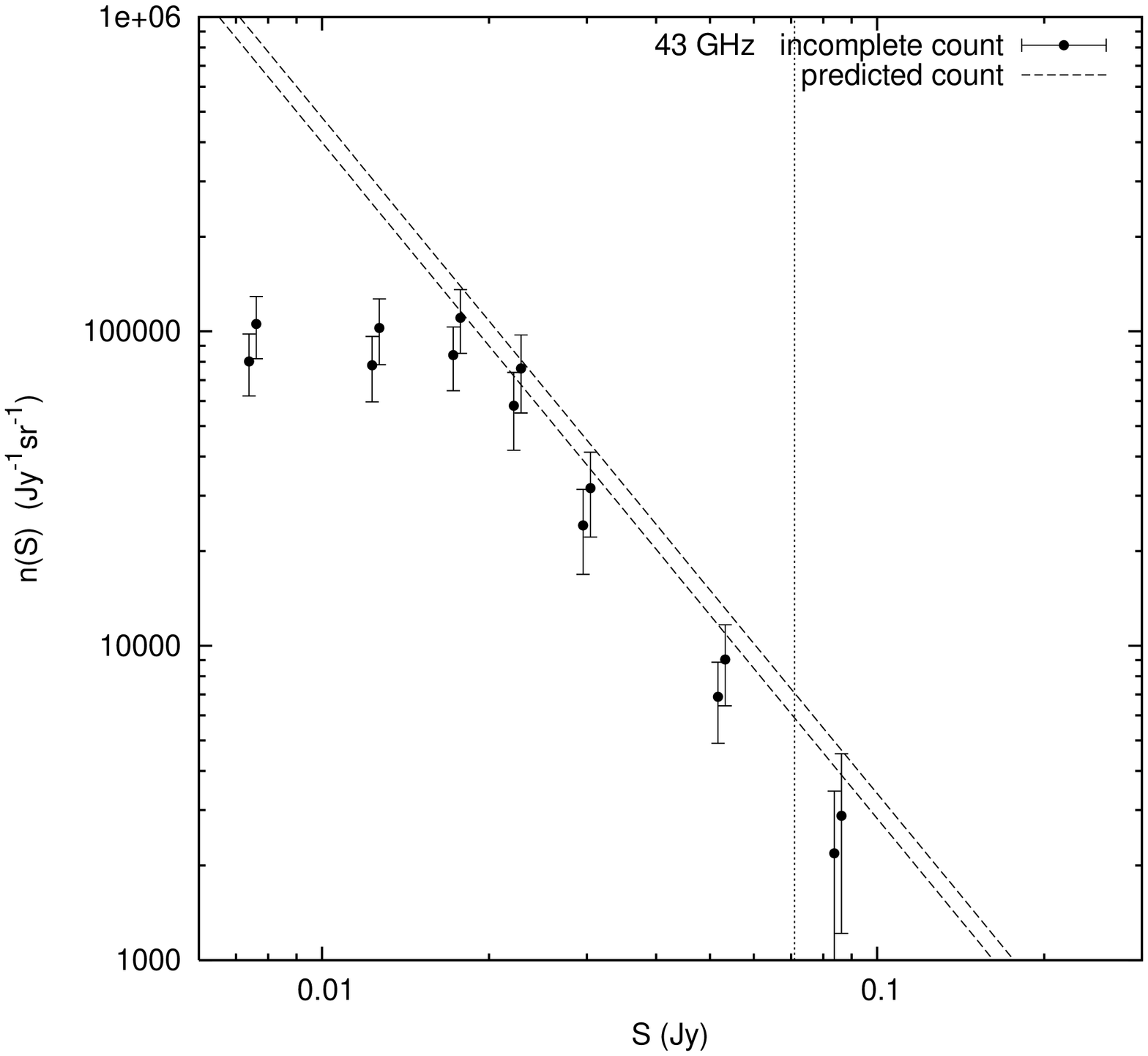,
        angle=270,width=8.0cm,clip=}} 
        \caption{The predicted counts for 22 and 43 GHz compared with the incomplete counts derived directly from the sample. The two dashed lines correspond to the uncertainty in $A$ from Table~2. The pairs of data points indicate the uncertainty in the area (111 to 145~deg$^{2}$) and the error bars are the Poisson errors. For clarity we have artificially separated the points within the pairs so that they are slightly above and below the corresponding value of $S$. The vertical dotted lines show the values of $S$ above which the direct counts are expected to be complete.}
\end{figure*}

\section{Error estimates}

There are a number of factors which contribute to the uncertainty in our values in $K$ and $A$. As well as the error in the mean of the distribution $k_{i} = r_{i}^{\rm 1-b}$, ($\sigma/\sqrt{m}$), there is also the error in the original 15~GHz source count: i.e. for $\rm{b} = 2.15 \pm {0.06}$ the error in the prefactor ($A$ at 15~GHz) at the centre of the data is $\sim5\%$. Combining these gives the uncertainties in $A$ shown in Table 5. 

  These are the \textit{minimum} estimates for the uncertainties. The errors hardest to quantify in our source count predictions lie in the assumption that the spectral index distribution is independent of flux density. We have seen from Table 2 and Figure 3 that we cannot detect a dependence within our current sample. Similarly, in Tables 3 and 4, the differences in the $K$ values for sources with $S_{15}~<~43$~mJy and those with $S_{15}~>~43$~mJy are not significant given the intrinsic errors. However, with only 6 sources above 100 mJy, our distributions of $k_{i}$ may be skewed towards values of $K$ more appropriate to the lower rather than the higher flux densities. Also, it is possible for sources outside our selected flux density range at 15 GHz, with a different spectral index distribution, to contribute to the predicted counts at another frequency, even within our estimated range of $S_{\rm min}$ to $S_{\rm max}$ for that frequency (see  section 11 for further discussion).

 We need to investigate how far the assumption of the independence of spectral index and flux density is a useful approximation for our present work. One way of testing our procedure has been to apply it to our data for 1.4 and 4.8~GHz, since we already have measurements of the source counts at these frequencies from earlier surveys (see section 8).

We can also gain some insight into the reliability of our method by comparing our predicted counts for 22 and 43 GHz with the incomplete counts derived directly from the sample, as in Figure 7. Here we show the two types of error on each data point: one is a systematic error due to the uncertainty in deriving the sample area from a fit to the known 15 GHz count (see section 2) and the other is the usual random Poisson error. We see that the predicted counts are close to the direct counts at the higher flux densities whereas at the lower flux densities there is a marked fall-off in the direct counts. This is as expected, since the original sample was complete to only 25 mJy. We have calculated the values of $S$ above which we might expect the direct counts to be complete, assuming an extreme rising spectral index, $\alpha_{15}^{22}$ or $\alpha_{15}^{43}$, of $-1$. These are 36~mJy for 22~GHz and 71~mJy for 43~GHz, and we see they are consistent with these plots.

\section{Lower frequency counts}

\begin{figure}
        {\epsfig{file=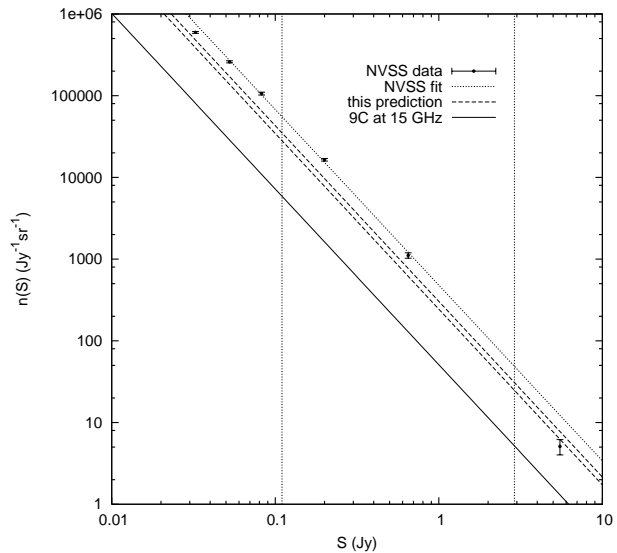,
        angle=270,width=8.0cm,clip=}} 
        \caption{Prediction for 1.4~GHz count, where the two dashed lines show the uncertainty in $A$ from Table 2. The vertical dotted lines indicate $S_{\rm min}$ and $S_{\rm max}$ }
\end{figure}

\begin{figure}
        {\epsfig{file=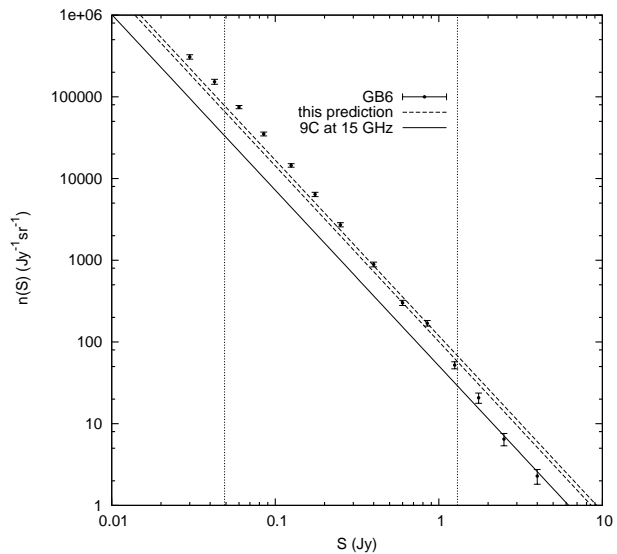,
        angle=270,width=8.0cm,clip=}} 
        \caption{Prediction for 4.8~GHz count, where the two dashed lines show the uncertainty in $A$ from Table 2. The vertical dotted lines indicate $S_{\rm min}$ and $S_{\rm max}$ }
\end{figure}

Figure 8 shows a plot comparing our predicted count at 1.4 GHz with a count from the NVSS survey (Condon et al. 1998). We have used 1.4 GHz data from the areas of our original 9C survey fields, as in Waldram et al. 2003. It can be seen that our prediction lies significantly below the data points in the range of flux density over which we might expect it to be valid. We estimate that an approximate fit to the data, keeping the count exponent at $-2.15$, is given by a value of $A$ of $\sim480$, as compared with the predicted value of $274\pm30$. This is not surprising because we already know that the assumption of the independence of spectral index and flux density does not hold over wide ranges of frequency and flux density, as can be seen, for example, in the 1.4 to 15~GHz spectral index distributions in Waldram et al. 2003 and Waldram \& Pooley 2004. In the latter paper we show that the percentage of inverted spectrum sources, with $\alpha_{1.4}^{15}<0$, increases with increasing flux density: we find that for three samples -- i.e. 5 to 25 mJy, 25 to 100 mJy, 100 mJy and above -- the percentages are 10, 20 and 33 respectively. In our current sample the percentage is also 20. This would suggest that below the completeness limit of our sample there is a source population with a higher proportion of steep spectrum sources and these  are consequently contributing an extra component to the count at 1.4~GHz.

We have also compared our 4.8~GHz prediction with the count from the Green Bank survey, using the data in Gregory et al. 1996, as illustrated in Figure 9. This corresponds to a frequency ratio of only $\sim3$, rather than $\sim10$, and we can see that there is agreement within the errors over a flux density range of approximately 0.175 to 1.25 Jy. At lower flux densities our predicted count is too low and at higher flux densities too high, which is consistent with the trend in spectral index found for 1.4~GHz.

\section{Higher frequency counts}

\begin{figure}
        {\epsfig{file=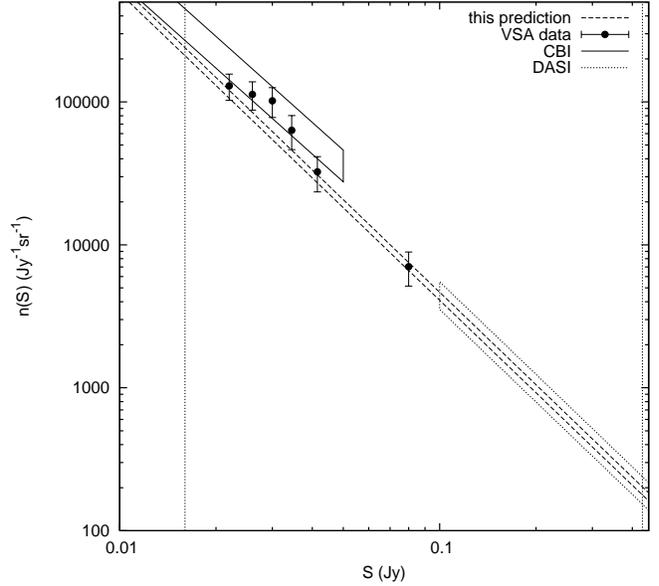,
        angle=270,width=8.5cm,clip=}} 
        \caption{Predicted 30 GHz count with VSA data and the measured CBI and DASI counts. The two dashed lines show the uncertainty in $A$ from Table 2. The vertical dotted lines indicate $S_{\rm min}$ and $S_{\rm max}$. }
\end{figure}

\begin{figure*}
        {\epsfig{file=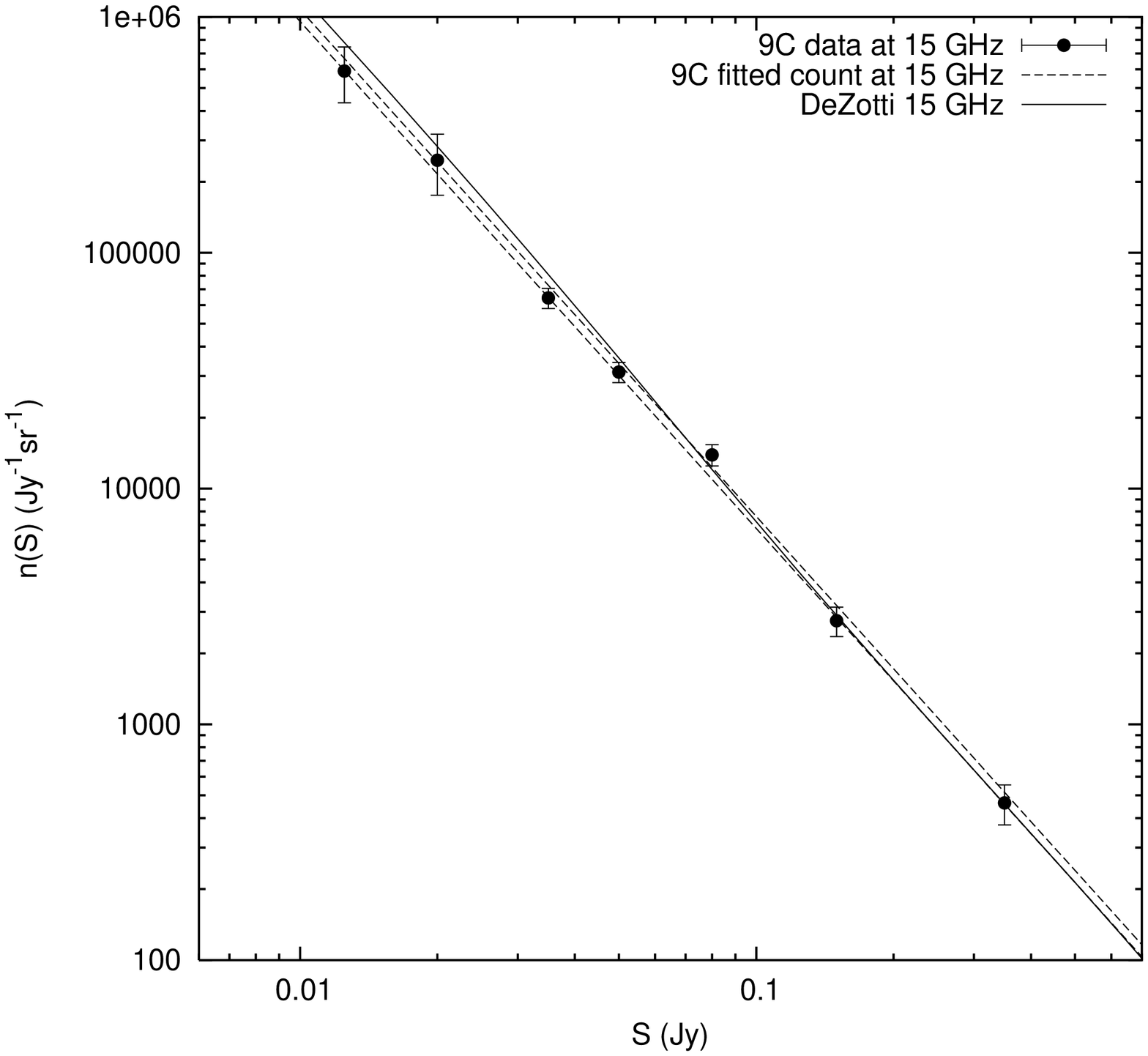,
        angle=270,width=8.0cm,clip=}} 
        {\epsfig{file=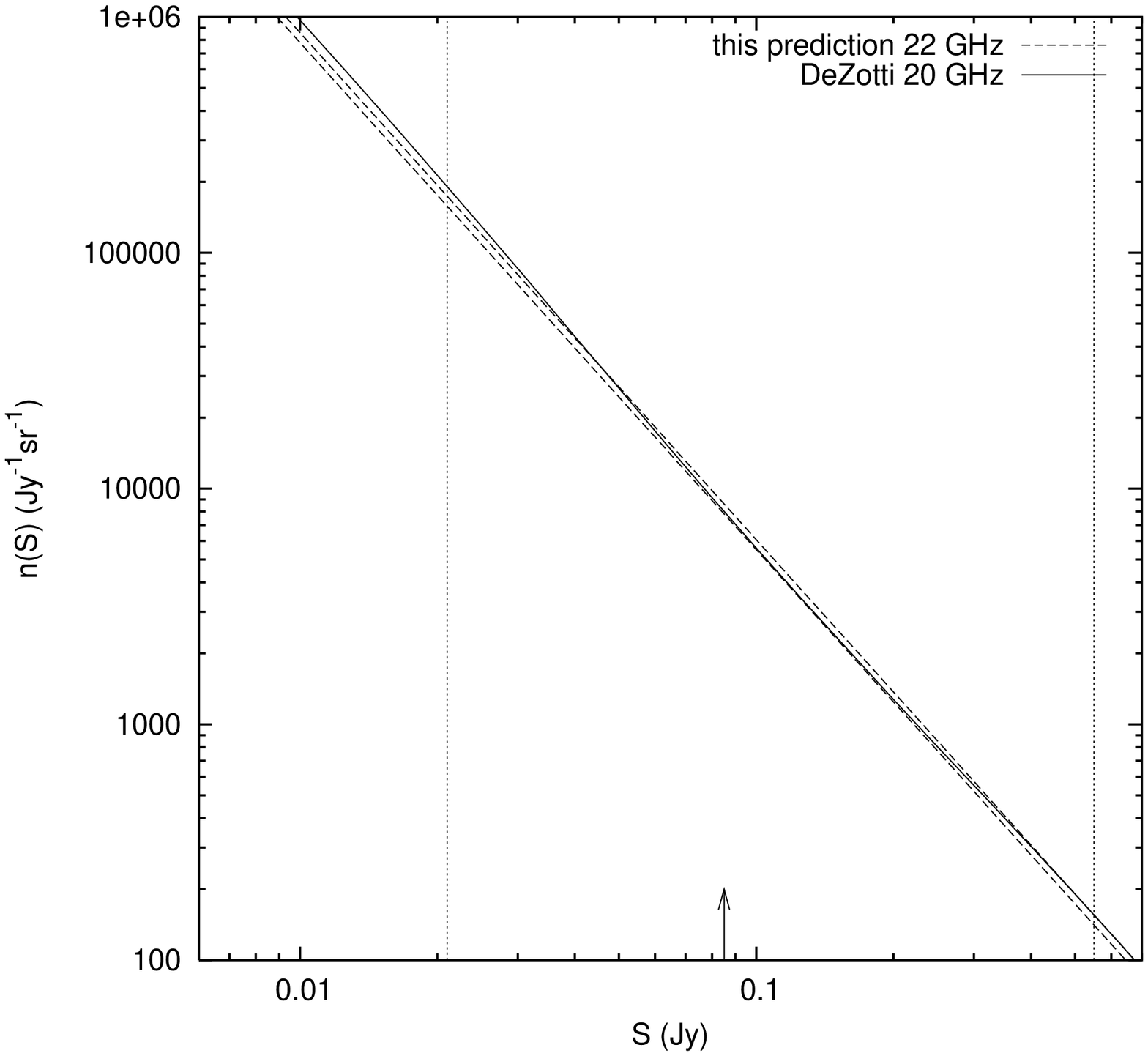,
        angle=270,width=8.0cm,clip=}} 
        {\epsfig{file=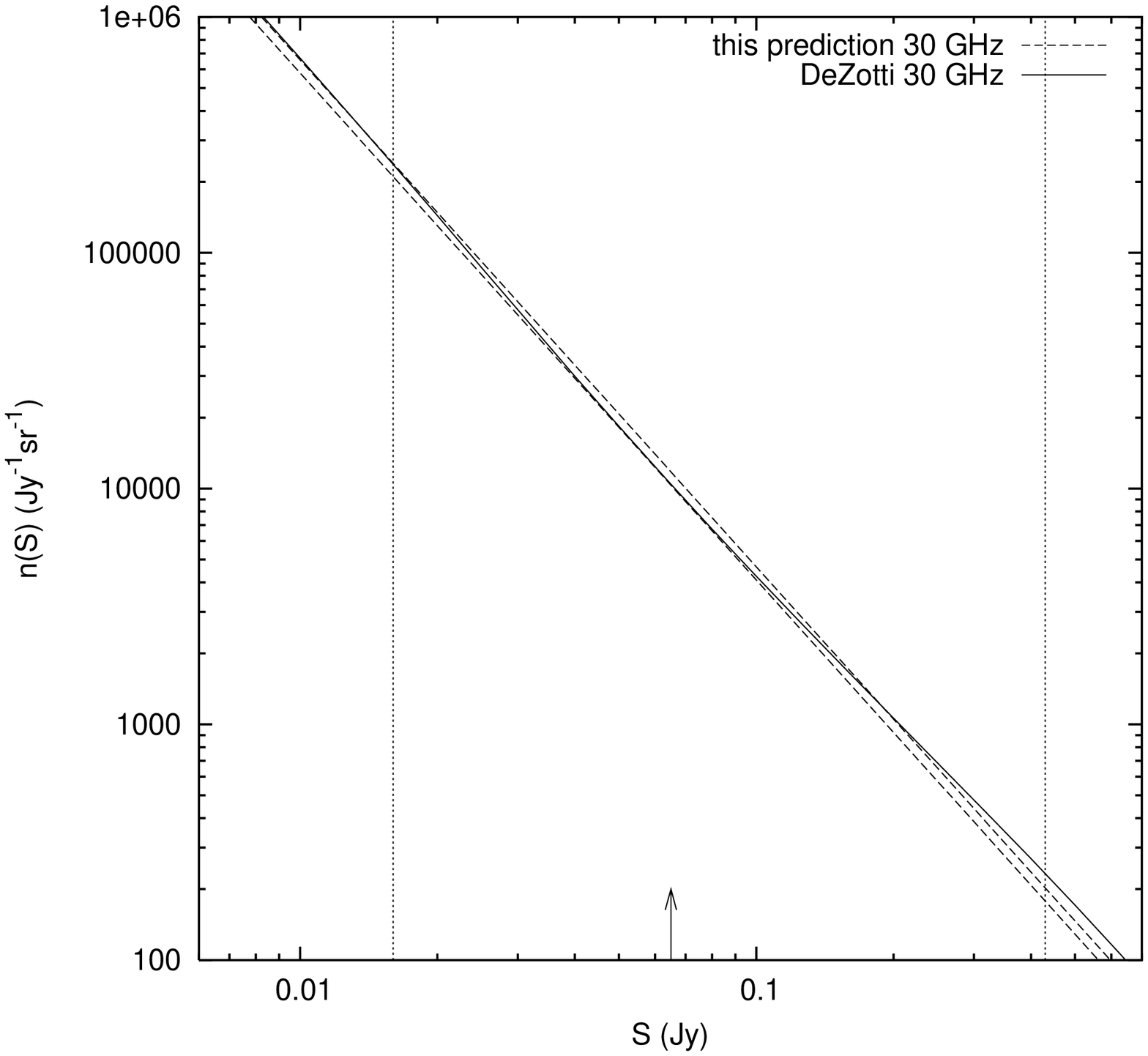,
        angle=270,width=8.0cm,clip=}} 
        {\epsfig{file=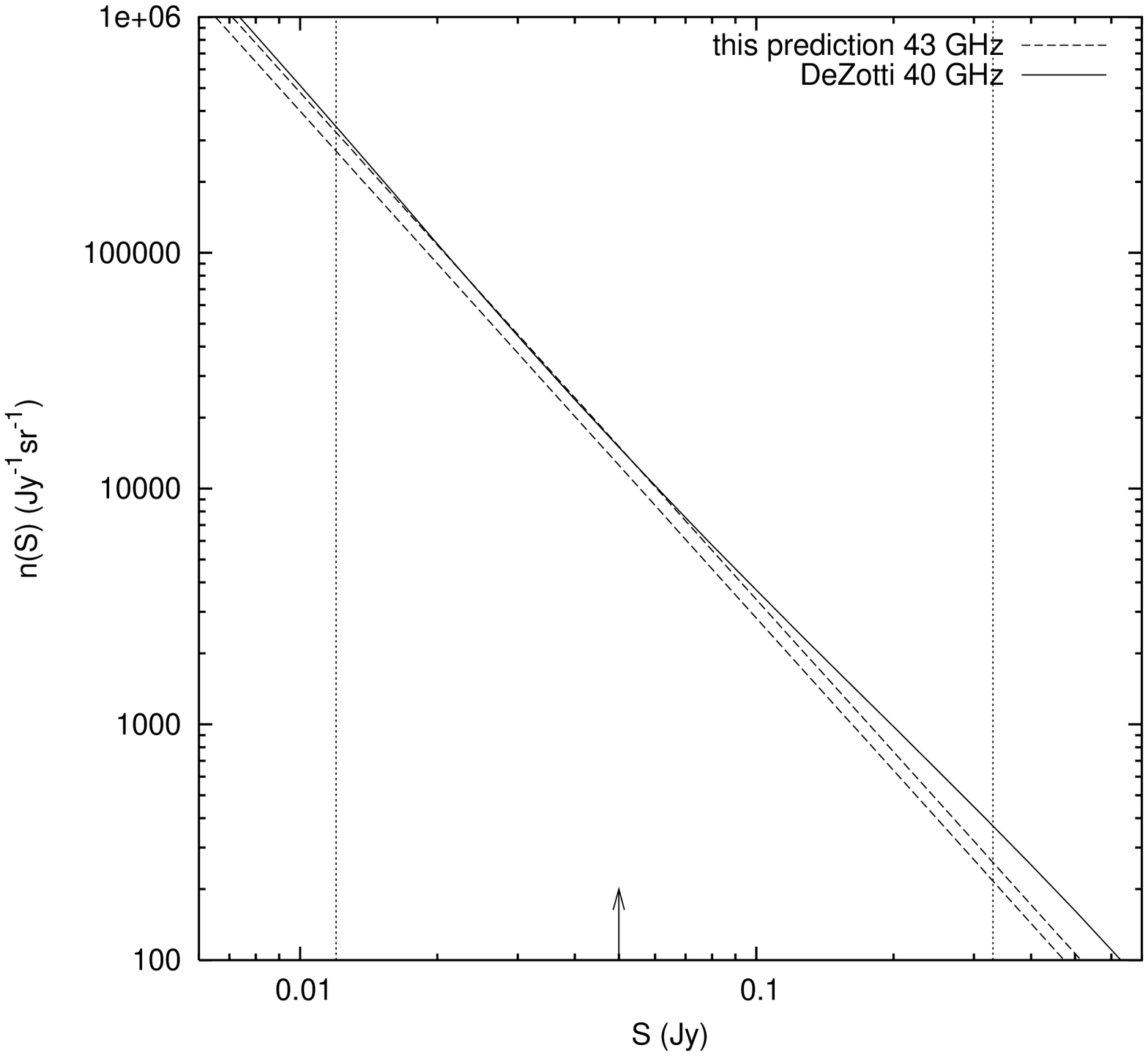,
        angle=270,width=8.0cm,clip=}} 
        {\epsfig{file=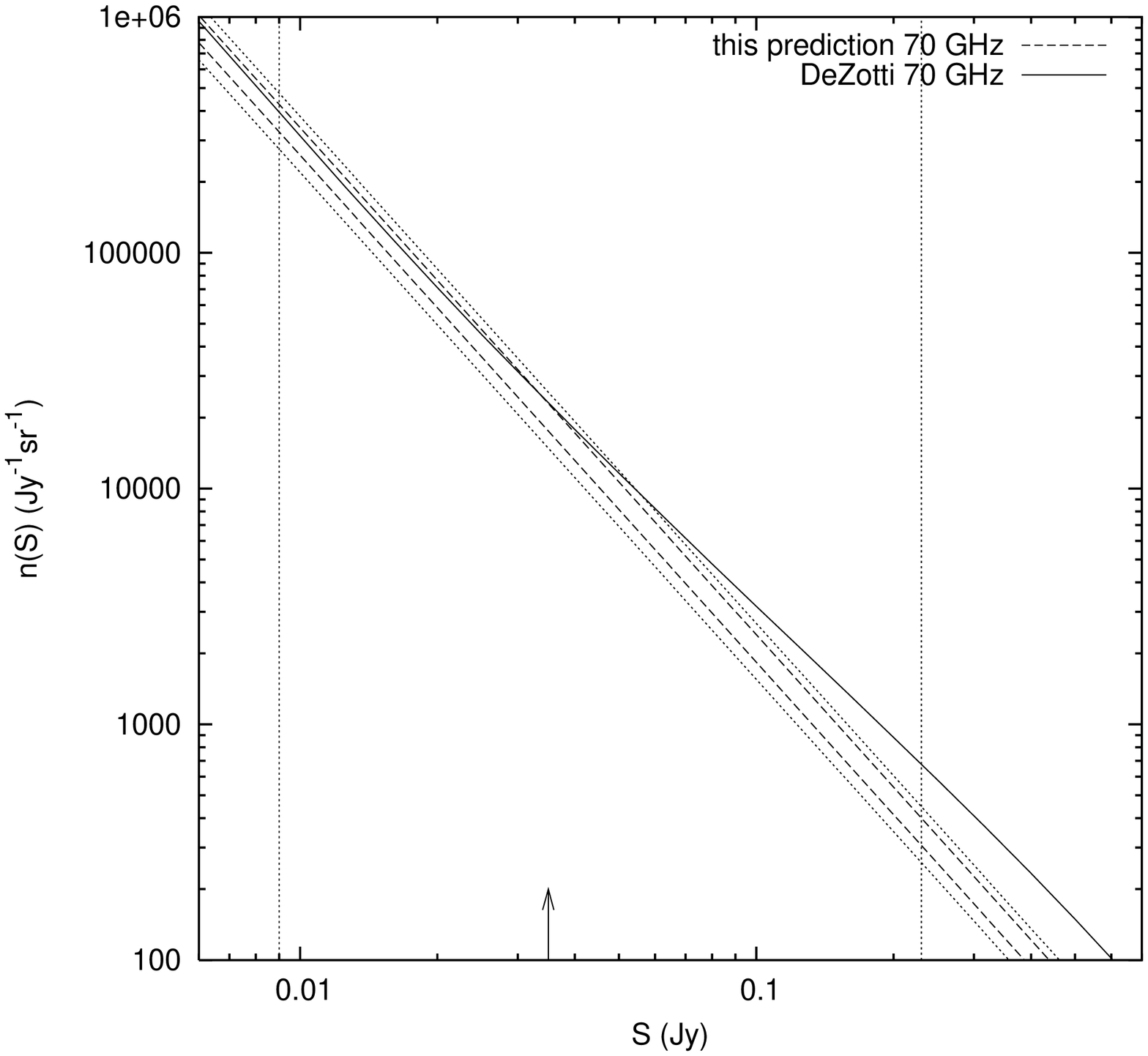,
        angle=270,width=8.0cm,clip=}} 
        {\epsfig{file=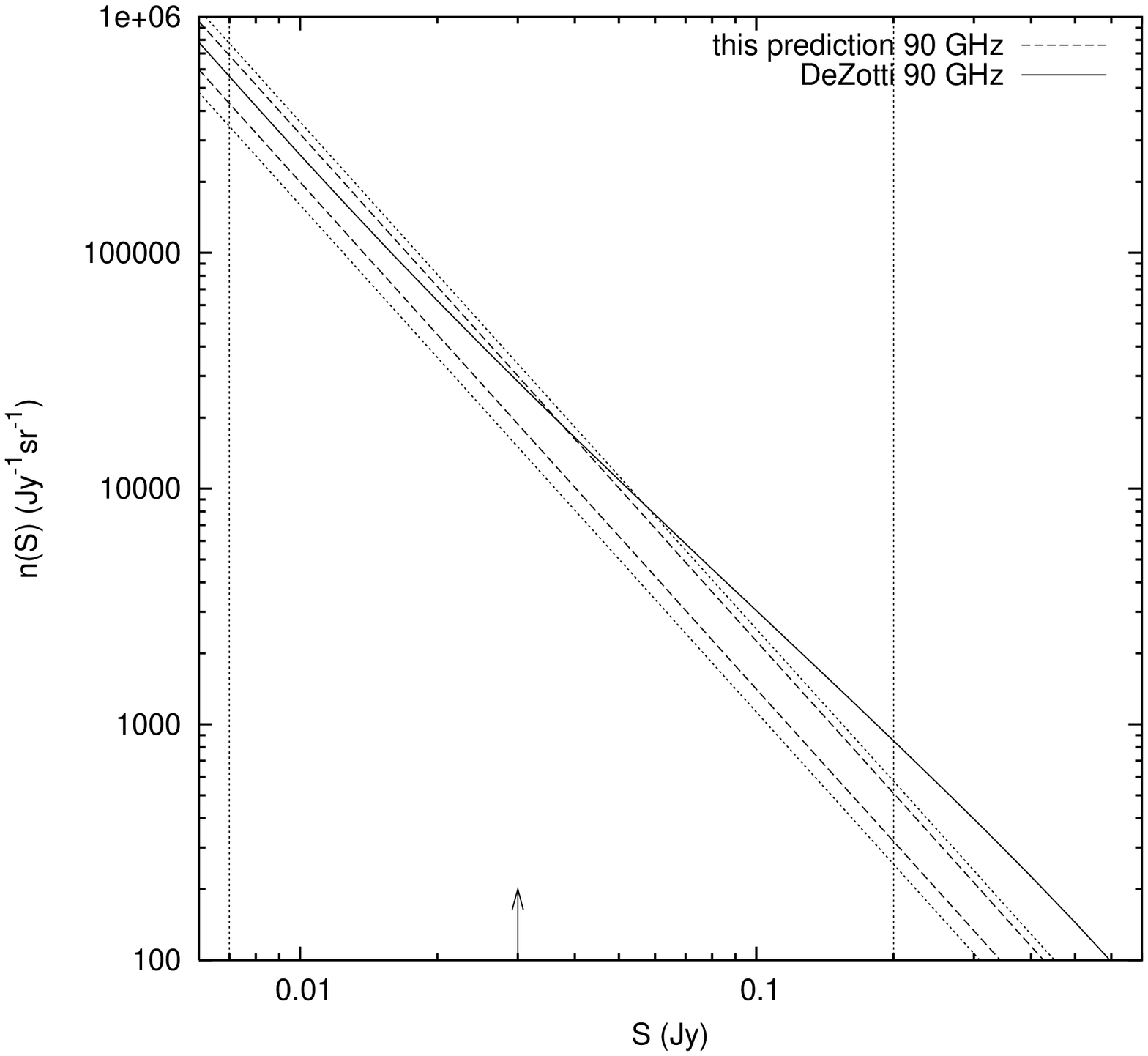,
        angle=270,width=8.0cm,clip=}} 
        \caption{Comparisons of our predictions, and also the known 9C 15~GHz count, with the de~Zotti models. The two dashed lines show the uncertainty in $A$ from Table 2. For 70 and 90~GHz these correspond to extrapolation of type (b), but the error limits for extrapolation of types (a) and (c) are also included (shown dotted). The vertical dotted lines indicate $S_{\rm min}$ and $S_{\rm max}$ and the vertical arrow marks the value of $S_{\rm c}$. }
\end{figure*}

At frequencies above 15 GHz the only data available with which to compare the predictions are those at 31~GHz from DASI (Kovac et al. 2002) and CBI (Mason et al. 2003) and at 33~GHz from the VSA (Cleary et al. 2005). Other source count data from ATCA (Ricci et al. 2004) or WMAP (Bennett et al, 2003), for example, lie outside the relevant flux density range. Figure 10 shows the comparison of our predicted count with the measured counts from DASI and CBI and with data from the VSA. \\ From DASI, over the range 0.1 to 10 Jy, we have taken \\
${\rm d}N/{\rm d}S_{31} = (32\pm7)(S_{31}/{\rm Jy})^{-2.15\pm0.20}\, {\rm Jy}^{-1}{\rm sr}^{-1}$ \\
and from CBI, over the range 0.005 to 0.05 Jy,\\
$N(>S_{31}) = (2.8\pm0.7)(S_{31}/10{\rm mJy})^{-1.0}\,{\rm deg}^{-2}$ \\
or \\
${\rm d}N/{\rm d}S_{31} = (92\pm23)(S_{31}/{\rm Jy})^{-2.0}\, {\rm Jy}^{-1}{\rm sr}^{-1}$ . \\
The VSA count, over the range 0.02 to 0.114 Jy, is fitted by \\
${\rm d}N/{\rm d}S_{33} = (21\pm4.5)(S_{33}/{\rm Jy})^{-2.34}\, {\rm Jy}^{-1}{\rm sr}^{-1}$ \\
We see from Figure 10 that for $S>0.1$~Jy our prediction is consistent with the DASI count and for $S<0.1$~Jy it is consistent with the VSA data, but it lies somewhat below the CBI count in the range $S<0.05$~Jy. 

Although no further direct data are available, we can compare our measured 9C count and our empirically predicted higher frequency counts with those from the models of de~Zotti et al. (2005); these plots are shown in Figure 11. Here the de Zotti models represent the sum of the contributions from the three main extragalactic source populations: FSRQs (flat spectrum radio quasars), BL Lacs (BL Lacertae type objects) and steep-spectrum radio sources. The contributions from other types of extragalactic source population are assumed to be negligible over the relevant ranges in flux density. We see that there is good agreement between our counts and the de Zotti models over the appropriate  $S_{\rm min}$ to $S_{\rm max}$ range for the frequencies 15, 20 and 30~GHz, but for 40, 70 and 90~GHz, although there is good agreement over the lower part of the range, below our upper `completeness' value $S_{\rm c}$, there is an increasing divergence at the higher flux densities, the models lying significantly above our counts.

In Figure 12 we have repeated the comparison of our 43~GHz prediction with the de~Zotti model, but here we have included the separate model components from  FSRQs, BL Lacs and steep spectrum sources. Taking appropriate tabulated values from the de~Zotti counts, we have made the following calculations. At a flux density of 316~mJy, a value close to the $S_{\rm max}$ of 330~mJy, our count is only  $65\pm6$~\% of the de~Zotti 40~GHz total count. At this point the de~Zotti model is dominated by the contribution from FSRQs, which amounts to 83~\% of the total. It is apparent that our empirical approach is predicting significantly fewer flat spectrum sources than the de~Zotti model in this higher flux density range. 

However, as we have seen, our 15 GHz sample contains few sources above 100~mJy and so our predicted count at another frequency is less reliable in the range $S_{\rm c}$ to $S_{\rm max}$.  It is possible that with increasing flux density there is a significant shift in the spectral index distribution towards flatter spectra, even if such a trend is not detectable in our data.

\begin{figure*}
        {\epsfig{file=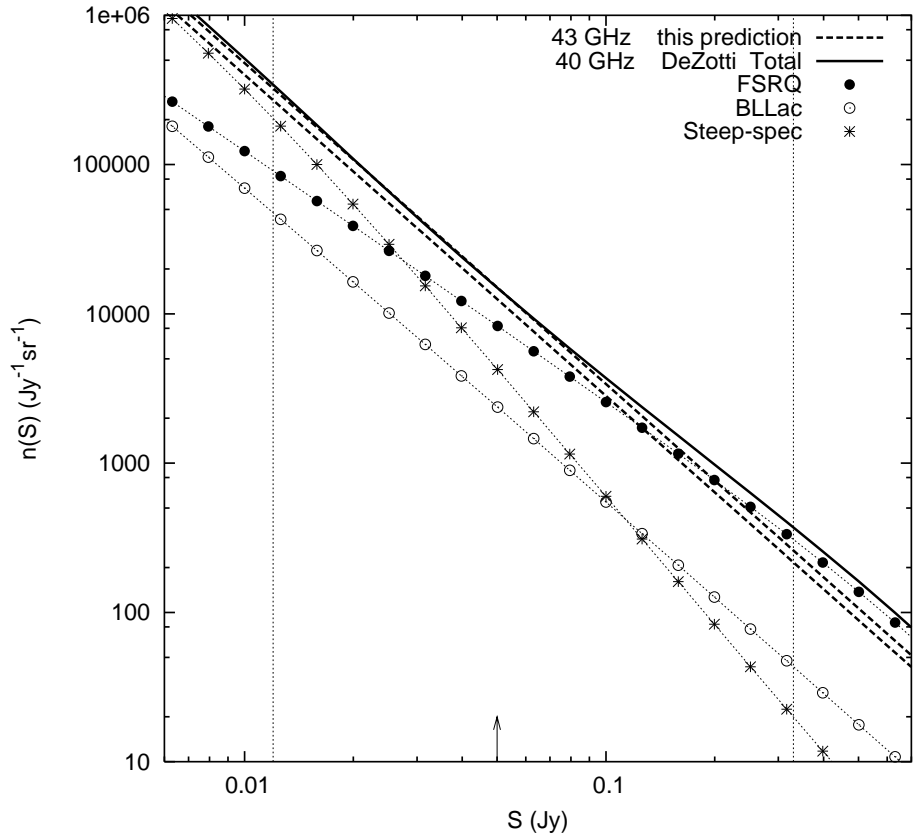,
        angle=270,width=12.0cm,clip=}}
        \caption{Comparison of our 43~GHz prediction with the de Zotti model, showing the separate model components from  FSRQs, BL Lacs and steep spectrum sources. The two dashed lines show the uncertainty in $A$ from Table 2. The vertical dotted lines indicate $S_{\rm min}$ and $S_{\rm max}$ and the vertical arrow marks the value of $S_{\rm c}$. }
\end{figure*}

\begin{table*}
\caption{Table showing the results of our tests for the effect of possible bias in the spectral index distributions. `FS' denotes flat spectrum sources ($\alpha_{15}^{\nu} < 0.5$) and `SS' denotes steep spectrum sources ($\alpha_{15}^{\nu} > 0.5$). }
\begin{tabular}{ccccccccc}

\hline
Frequency & \multicolumn{3}{c}{point sources} & \multicolumn{3}{c}{resolved sources} & $K$ & $K$ \\ 
GHz & no. of FS & no. of SS & mean $k_{\rm i}$ for SS & no. of FS & no. of SS & mean $k_{\rm i}$ for SS & test value & value used \\
\hline

22.0 & 40 & 33 & $0.66\pm0.02$ & 10 & 27 & $0.62\pm0.03$ & $0.81\pm0.02$ & $0.80\pm0.02$ \\
43.0 & 27 & 46 & $0.30\pm0.02$ &  4 & 33 & $0.23\pm0.02$ & $0.46\pm0.03$ & $0.44\pm0.04$ \\

\hline
\end{tabular}
\end{table*}

\section{Phase calibrators for ALMA at 90 GHz}
One strategy proposed for the ALMA phase calibration is to use point sources with flux densities above 20 mJy at 90 GHz and to extrapolate the phase solutions up to the appropriate target frequency (see, for example, Holdaway \& Owen, 2005). It is therefore important to investigate whether there will be a sufficient density of such sources available at this frequency. Holdaway \& Owen have developed a simple parametrized model of the source population, and, using observed 8.4 GHz and 90 GHz fluxes, estimated the source counts as a function of frequency.  They have estimated a count of about 1800 point sources per steradian brighter than 20~mJy at 90~GHz.

As far as our own 90~GHz prediction is concerned, our estimate of the number of sources in the range 20 to 200~mJy (where 200~mJy is the value of $S_{\rm max}$) is $940\pm220$~${\rm sr}^{-1}$. However, not all of these sources will be of a  sufficiently small angular size for use as calibrators. At 15~GHz we should expect about half the sources to be less than 0.1 arcsec in angular diameter (see paper~2), though at 90~GHz it is likely to be a somewhat higher fraction. Our estimate is thus equivalent to approximately one such source in every 6 or 7~square degrees in this flux density range.

If this is correct (though it may well be an underestimate), it suggests that suitable phase calibrator sources for ALMA may be as much as twice as far from the target source as has been assumed hitherto, meaning that either longer slews or longer on-source integrations will be required to achieve good phase solutions.  However, given the very high sensitivity of ALMA, this is not anticipated to affect significantly the observing efficiency.

\section{Discussion}

As we have seen, our empirical method of predicting the source counts depends on two main assumptions: first, that our sample provides a typical distribution of spectral indices and secondly, that this distribution is independent of flux density.

We consider first the reliability of our spectral index measurements. Since sources at these radio frequencies can be extremely variable (see paper 3) it was essential to make our follow-up observations simultaneously and the fact that we were able to do so is an important element in our method. However, it has meant that, at any one time, the VLA measurements were made with the same configuration of the telescope for all frequencies, leading to a wide variation in the size of the synthesized beam across the frequency range (see paper 1). Although we used integrated flux densities at all frequencies, it is possible that some flux of the more extended sources has been `resolved out' at the higher frequencies, resulting in a bias towards steeper spectra.

To investigate the effect this might have on our predictions, we divided the sample into two groups, those sources which appeared point-like at all frequencies (73) and those which were resolved at one or more frequencies (37). For each of the frequencies 22 and 43~GHz, we further divided the groups into `FS' or flat spectrum sources ($\alpha_{15}^{\nu} < 0.5$) and `SS' or steep spectrum sources ($\alpha_{15}^{\nu} > 0.5$). Our results are shown in Table 6. We see that at both frequencies the mean value $k_{\rm i}$ for the resolved SS sources is lower than the value for the point SS sources and that at 43 GHz the difference appears to be marginally significant. Of course, we cannot tell whether this means that the spectra of the resolved SS sources are genuinely steeper, which is quite possible, or whether we are actually missing flux from over-resolution. We can, however, assign the mean $k_{\rm i}$ value of the point SS sources to each of the resolved SS sources, and calculate a test value of $K$ for the whole sample  for comparison with the $K$ value used in our predictions. We find (see Table 6) that these do not differ significantly, meaning that, even if we were losing flux of some of the extended sources, the error is likely to be small.

This conclusion is corroborated by the fact that the predicted counts at 30~GHz, calculated by interpolating the source spectra between 22 and 43~GHz, are in good agreement with experiment.

We have seen from section 3 that our second main assumption, that the spectral index distribution is independent of flux density, leads to a form of the predicted count with the same exponent b as at 15~GHz but a different pre-factor $A$. Any variation in the spectral index distribution with flux density can be envisaged as resulting in a dependence of $A$ on flux density, equivalent to curvature in the logarithmic count.   

We have already emphasized earlier in the paper that our data are sparse at the higher flux densities. Thus, although we have shown that, within our sample, spectral index and flux density appear to be independent, we cannot assume that this continues to hold above a 15~GHz flux density of $\sim100$~mJy. We have, though, been able to assemble from a wider area a small un-biassed sample of 16 sources (including 5 from the original sample) with 15~GHz flux densities in the range 102 to 784~mJy and simultaneous follow-up observations at 15, 22 and 43~GHz. We find that they all have spectral indices $\alpha_{22}^{43}$ in the range 0.1 to 1.1, indicating a possible flattening of the spectra compared with the values shown in Figure~4. A calculation of the value of $A$ at 43~GHz from these 16 sources gives $A = 27\pm2$, as compared with $22\pm2$ in Table 5, corresponding to a count at 316~mJy of $79\pm7$~\% of the de~Zotti 40~GHz total count, as compared with the $65\pm6$~\% quoted in section 9. There is thus some evidence for an increase in the value of $A$ and consequently for closer agreement with the de~Zotti prediction; however, given the error estimates, it is scarcely conclusive. It is clear that more data at these higher flux densities would be required to detect any significant change in the slope of the count.

\section{Conclusions}
We have shown that it is possible to use our multi-frequency follow-up observations of a sample of sources from the 9C survey at 15~GHz to make some empirical estimates of the source counts at higher radio frequencies.  These predictions are important, in spite of the necessary limitations of our method, since at present there are few direct observations at these frequencies. Our data, although indirect, have two particular advantages: the measurements for any one source were made simultaneously, thereby avoiding problems with variability which can be extreme in some cases, and they also extended over a wide range of frequency, reaching as high as 43~GHz. We find our results to be consistent with the known counts at 30~GHz and in good agreement with the models of de~Zotti et al. (2005) below 43~GHz; but for frequencies of 43~GHz and above, although there is agreement at the lower flux densities, our counts diverge progressively from those of de~Zotti at the higher values, in that our predictions imply significantly fewer flat spectrum sources. However, our data are sparse above a 15~GHz flux density of $\sim100$~mJy and we cannot rule out the possibility  that with increasing flux density there is a significant shift in the spectral index distribution towards flatter spectra, although this is not detected with any certainty in our measurements.

The forthcoming wide-area survey with the Australia telescope (AT20G) should provide definitive source counts in the higher flux density range at a frequency of 20~GHz. Our own work can be seen as complementary in that it is applicable to somewhat lower flux densities and higher frequencies.

\section*{Acknowledgments}
We are grateful to the staff of our observatory for the operation of the Ryle Telescope, which is funded by PPARC. We also thank Gianfranco de Zotti for providing us with his modelled counts in numerical form and John Richer for discussion on ALMA calibration.

\label{lastpage}

\end{document}